\documentstyle[12pt,ssi,epsfig]{article}
\newcommand{\klmm}    {\mbox{$K^\circ_L \! \rightarrow \! \mu^+ \mu^-$}}
\newcommand{\klee}    {\mbox{$K^\circ_L \! \rightarrow \! e^+ e^-$}}
\newcommand{\kleeg}   {\mbox{$K^\circ_L \! \rightarrow \! e^+ e^-\gamma$}}
\newcommand{\kleegg}  {\mbox{$K^\circ_L \! \rightarrow \! e^+ e^-\gamma\gamma$}}
\newcommand{\kleeee}  {\mbox{$K^\circ_L \! \rightarrow \! e^+ e^- e^+ e^-$}}
\newcommand{\klme}    {\mbox{$K^\circ_L \! \rightarrow \! \mu e$}}
\newcommand{\klpee}   {\mbox{$K^\circ_L \! \rightarrow \! \pi^\circ e^+ e^-$}}
\newcommand{\klpnn}   {\mbox{$K^\circ_L \! \rightarrow \! \pi^\circ \nu \overline{\nu}$}}
\newcommand{\klpp}    {\mbox{$K^\circ_L \! \rightarrow \! \pi^\circ \pi^\circ$}}
\newcommand{\klpen}   {\mbox{$K^\circ_L \! \rightarrow \! \pi^\pm e^\mp \nu_e$}}
\newcommand{\kmn}     {\mbox{$K^+ \! \rightarrow \! \mu^+ \nu_\mu $}}
\newcommand{\kpp}     {\mbox{$K^+ \! \rightarrow \! \pi^+ \pi^\circ $}}
\newcommand{\kmuiii}  {\mbox{$K^+ \! \rightarrow \! \pi^\circ \mu^+ \nu_\mu $}}
\newcommand{\kpen}    {\mbox{$K^+ \! \rightarrow \! \pi^\circ e^+ \nu_e $}}
\newcommand{\kpmn}    {\mbox{$K^+ \! \rightarrow \! \pi^\circ \mu^+ \nu_\mu $}}
\newcommand{\kppen}   {\mbox{$K^+ \! \rightarrow \! \pi^+ \pi^- e^+ \nu_e $}}
\newcommand{\kmng}    {\mbox{$K^+ \! \rightarrow \! \mu^+ \nu_\mu \gamma$}}
\newcommand{\sdp}     {\mbox{${\rm SD}^+$}}
\newcommand{\kmnmm}   {\mbox{$K^+ \! \rightarrow \! \mu^+ \nu_\mu \mu^+ \mu^-$}}
\newcommand{\kmnee}   {\mbox{$K^+ \! \rightarrow \! \mu^+ \nu e^+ e^-$}}
\newcommand{\kenee}   {\mbox{$K^+ \! \rightarrow \! e^+ \nu e^+ e^-$}}
\newcommand{\kpnn}    {\mbox{$K^+ \! \rightarrow \! \pi^+ \nu \overline{\nu}$}}
\newcommand{\kpX}     {\mbox{$K^+ \! \rightarrow \! \pi^+ X^\circ$}}
\newcommand{\kpgg}    {\mbox{$K^+ \! \rightarrow \! \pi^+ \gamma \gamma$}}
\newcommand{\kpee}    {\mbox{$K^+ \! \rightarrow \! \pi^+ e^+ e^-$}}
\newcommand{\kpmm}    {\mbox{$K^+ \! \rightarrow \! \pi^+ \mu^+ \mu^-$}}
\newcommand{\kpme}    {\mbox{$K^+ \! \rightarrow \! \pi^+ \mu^+ e^-$}}
\newcommand{\pee}     {\mbox{$\pi^\circ \! \rightarrow \!  e^+ e^-$}}
\newcommand{\pgX}     {\mbox{$\pi^\circ \! \rightarrow \!  \gamma X^\circ$}}
\newcommand{\pnn}     {\mbox{$\pi^\circ \! \rightarrow \!   \nu \overline{\nu}$}}
\newcommand{\pme}     {\mbox{$\pi^\circ \! \rightarrow \!  \mu^+ e^-$}}
\newcommand{\vtd}     {\mbox{$|V_{td}|$}}
\newcommand{\vus}     {\mbox{$|V_{us}|$}}
\newcommand{\XPT}    {\mbox{CHPT}}
\begin{document}

\begin{tabular}{cccccr}
\hspace{1.2cm} & \hspace{1.2cm} & \hspace{1.2cm} & \hspace{1.2cm} & \hspace{1.2cm} &\
  BNL Preprint {\bf \underline{65021}}\\
 & & & & & hep-ex/9801016\\
 & & & & & January 16, 1998\\
\end{tabular}

\title{Rare and Forbidden Kaon Decays at the AGS.}

\author{Steve Kettell\thanks{Supported by DOE Contract DE-AC02-76CH00016. 
\vskip 0.5in 
\noindent
\copyright\ 1997 by Steve Kettell. }\\ 
Brookhaven National Laboratory \\
Upton, NY 11973 \\[0.4cm]
}

\maketitle

\begin{abstract} An overview of the Rare Kaon Decay program at the AGS is presented,
with particular emphasis on the three major experiments currently running and analyzing
data. A brief
overview of earlier kaon decay experiments and recent improvements in AGS performance is also provided. 
This review concludes with a discussion
of proposed and developing experiments planned to run in the year 2000 and beyond (AGS---2000). % concludes this review.
\end{abstract}

\section{Overview}

The study of kaon decays has had a long and rich history at the AGS, including a Nobel
Prize for the discovery of CP violation in the K$_L$ decay to two pions~\cite{pipi} in 1963
and the observation of the suppression of Flavor Changing Neutral Currents (FCNC) in
the \klmm\ decay.~\cite{carithers}
The study of rare kaon decays %XXXhas 
had a renaissance in the early 1980's that has continued
up to the present. The third generation of these experiments is currently running.

\begin{figure}[htb]
\begin{minipage}[htb]{.44\linewidth}
\centering\epsfig{file=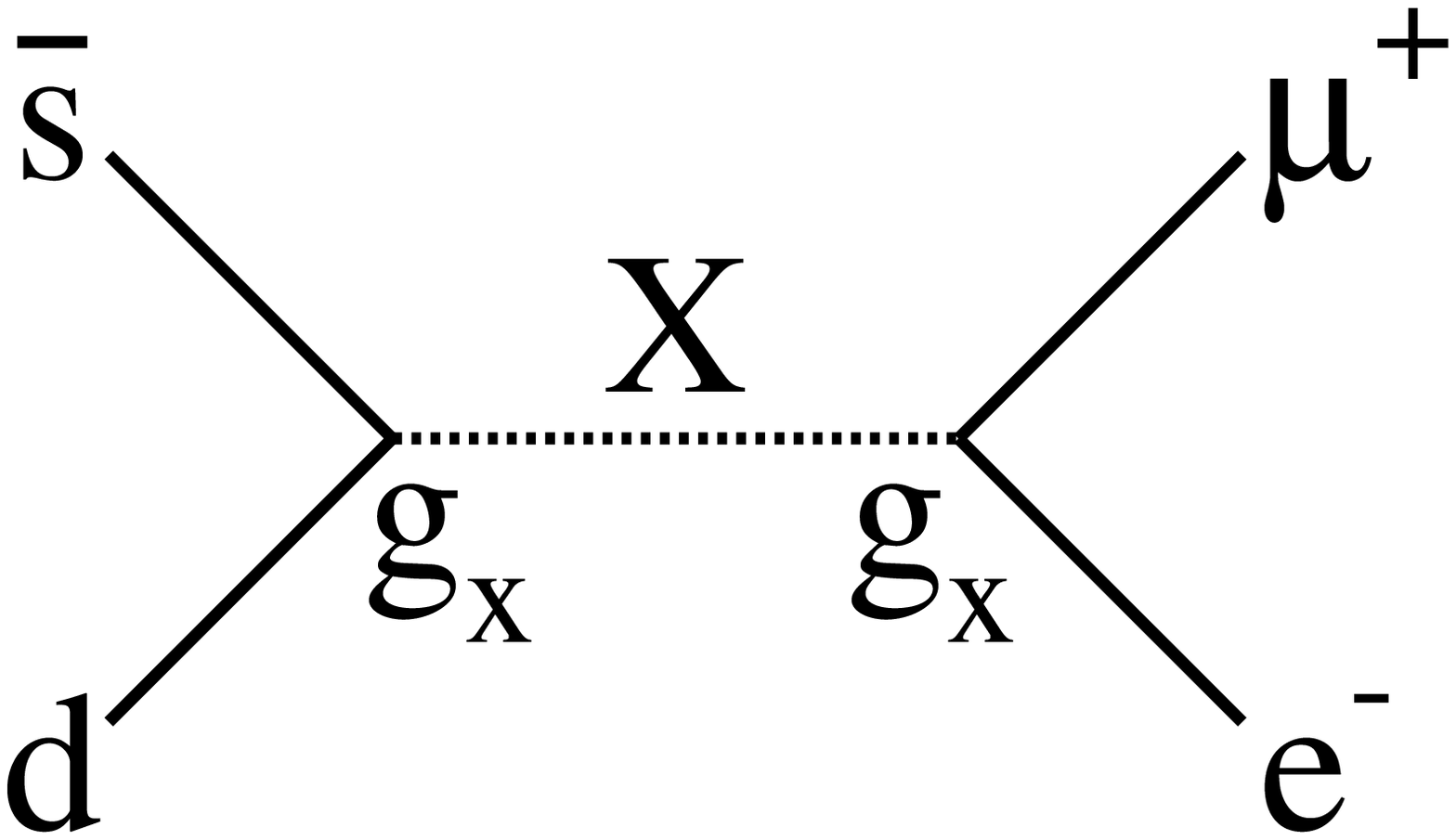,width=\linewidth,angle=0}
\end{minipage}\hfil 
\begin{minipage}[htb]{.44\linewidth}\vspace{-2.7cm}
\centering\epsfig{file=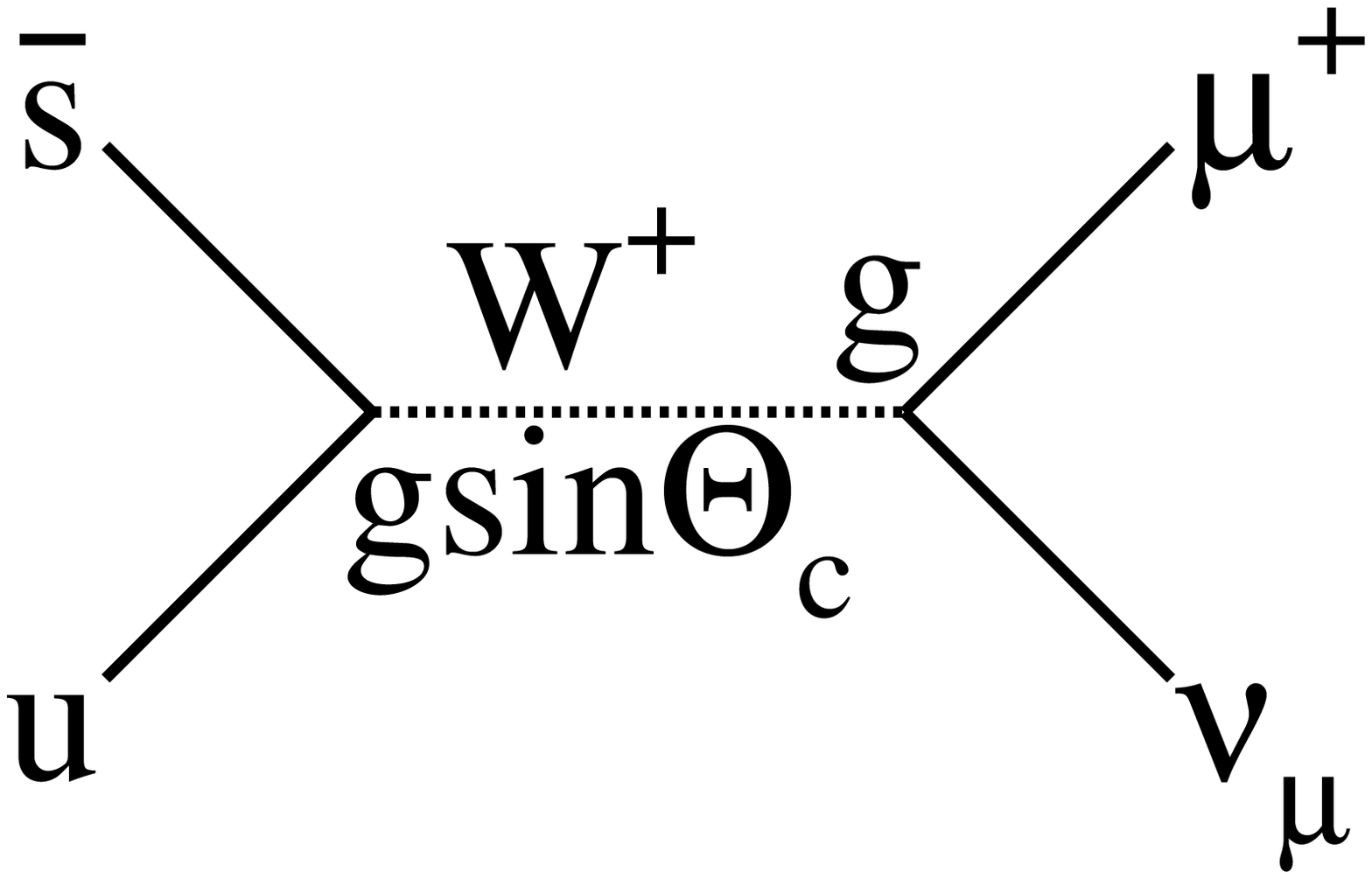,width=\linewidth,angle=0}
\end{minipage}
{\caption{\label{LFV} Heavy gauge boson mediating \klme, and $W^+$ mediating \kmn.}}
\end{figure}

This renaissance was motivated primarily by the realization that with the large numbers
of kaons available at the AGS and with modern detectors and data acquisition apparatus, a tremendous leap
forward in sensitivity was possible. At the same time it was realized that at such 
sensitivities the reach in mass scale for possible new interactions of these experiments
is quite large indeed.~\cite{ritchie,buchholz} For example, a comparison of the lepton flavor violating decay 
\klme\ with ordinary \kmn\ decay
(see figure~\ref{LFV}), leads to an expression for
the mass ($M_X$) for a new gauge boson X (assuming the same V-A interaction), 
in terms of the \klme\ branching ratio of:
\begin{equation}
M_X \approx 220 TeV/c^2 \Biggl[{g_X \over{g}}\biggr]^{1/4} \Biggl[ {10^{-12} \over{B(K_L \to 
\mu e)}} \Biggr]^{1/4}
\label{eqn:mue_munu}
\end{equation}
For $g_X\sim g$ the mass scale reached is far beyond that of any planned accelerator.
Likewise, for a purely vector interaction, the companion decay \kpme\ can probe 
similarly high mass scales.
Lepton flavor violation is not forbidden by an underlying gauge symmetry,
so while it is not allowed in the Standard Model (SM), further experimental verification 
becomes even more compelling. Several proposed extensions to the SM do allow 
lepton flavor violation.~\cite{cahn80,ellis,desh,pati,langacker,mukh}

The huge suppression of the decay rates of FCNC by the 
GIM mechanism~\cite{GIM} leaves a large `window of opportunity' for discovery of non-SM
processes. 
In addition to the searches for physics explicitly beyond the standard model, 
significant increases in sensitivity for SM allowed decays can close the window
on non-SM processes, such as \klee\ or 
\kpnn~\cite{hagelin,grossman,couture,bigi,agashe,leurer,davidson,bertolini,wilczek}.

\begin{figure}[htb] 
\center\epsfig{file=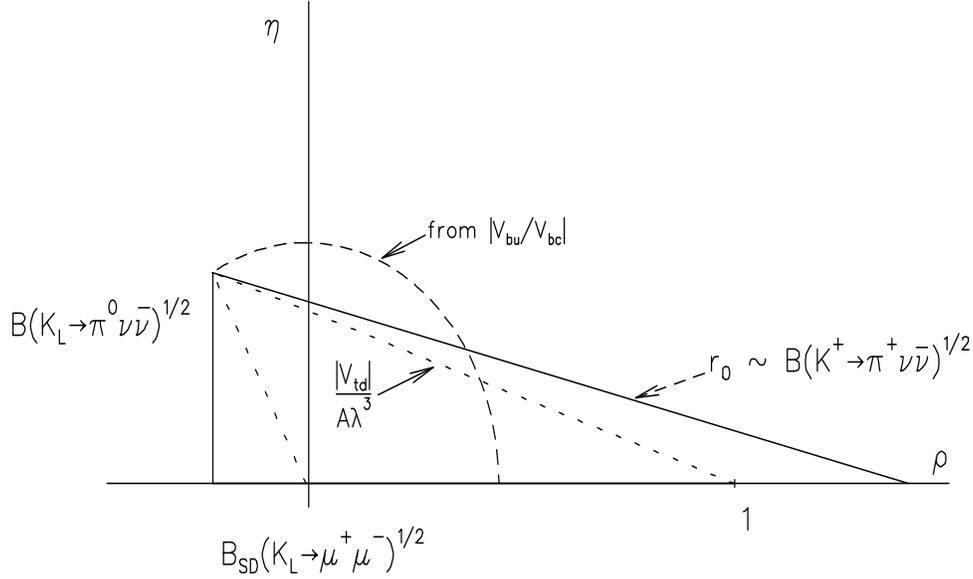,width=5.in,height=3.in,angle=0}
{\caption{\label{triangle} Unitarity triangle.} }
\end{figure}

Measurements of the branching ratios for the SM processes \kpnn\ or \klmm\ provide
means to determine fundamental SM parameters, such as the CKM matrix element
\vtd~\cite{CKM} and $\rho$~\cite{wolf}. A proposed
experiment to measure \klpnn\ would be able to determine the SM parameter $\eta$.~\cite{wolf}
From the kaon system alone a complete determination of the `unitarity triangle'
can be made (see figure~\ref{triangle}). The interpretation of \klmm\ is complicated by the presence of long
distance effects, but both \kpnn\ and \klpnn\ give very clean determinations of the
fundamental parameters and the two of them together are enough to completely determine
the triangle.~\cite{bf,bbl}

Finally, several `modestly rare' decays have been seen, with branching ratios from
$10^{-5}$--$10^{-8}$. These are generally long-distance dominated and are difficult to calculate.
Many of these serve as a testing ground for low-energy effective-theories,
such as Chiral Perturbation Theory (\XPT).~\cite{XPT} Some examples include: \kpee, \kpmm, \kpgg, 
\kmng(\sdp), \kppen, \kmnee, and \kenee.
Some non-rare decays, such as \kpen, are being pursued to more accurately determine the
CKM matrix element \vus.
\section{AGS}

\begin{figure}[htb] %\vspace{-1.cm}
\center\epsfig{file=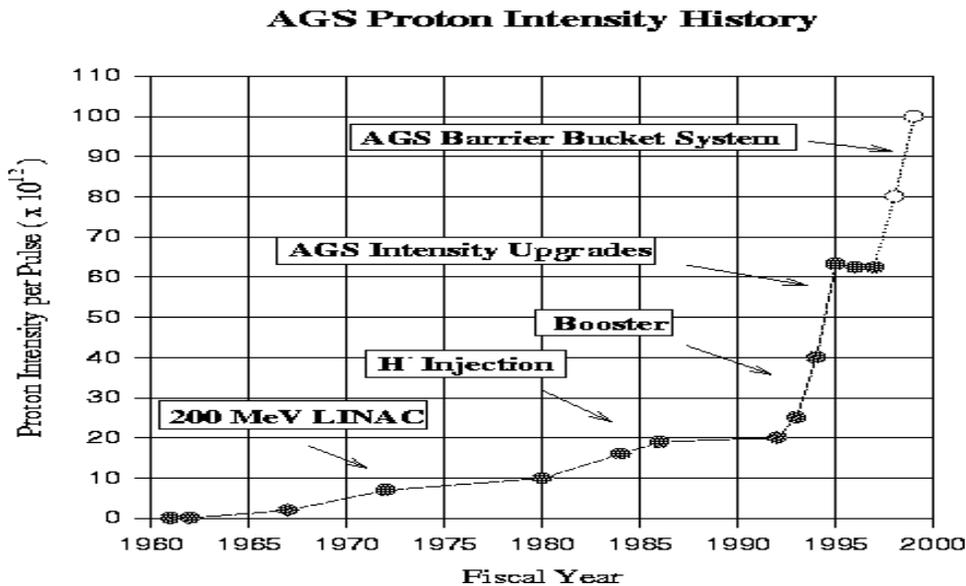,width=5.in,height=3.in,angle=0}
%\vspace{-1.5cm}
{\caption{\label{ags_ppp} AGS peak accelerated proton intensity per year.} }
\end{figure}

The AGS has made continued and significant increases in the number of protons accelerated in the
machine and available
for experiments (see Fig.~\ref{ags_ppp}).
%\vspace{-1.5cm}
%\vspace{-1.5cm}
 A very rapid increase over the last few years has occurred with the addition of the
AGS Booster. 
With the Booster (and several additional AGS upgrades)~\cite{brenan} the intensity of accelerated protons
has increased to $\sim$60 Tp (Tp $\equiv10^{12}$ protons/spill).
Further increases are expected with the addition of a RF Barrier Bucket during the 1998--99
running cycle. This Barrier Bucket should allow the stacking of additional Booster pulses with
the possibility of reaching the AGS space charge limit of $\sim$150 Tp.

A large number of other improvements to the operation of the AGS have been achieved. 
The Duty Factor (DF) has increased both macro- and micro-scopically. The macro DF has been increased from 30\% to 44\%
by increasing the spill length and reducing the inter-spill period. There are plans to increase
this even further.  Additional work on reducing ripple in the spill has increased the effective
length of the spill, even at the longer spill length. The AGS is also able to switch back and forth
between Fast Extracted Beam (FEB) and Slow Extracted Beam (SEB) on a pulse by pulse basis. This 
allows the kaon program (using SEB) to run simultaneously with the tune-up for experiments such as the
muon g-2 (using FEB). Studies have also demonstrated the ability to extract SEB with  bunching of 
the extracted beam.~\cite{glenn} This is useful for experiments that want to do time-of-flight for low-energy
$K_L^\circ$'s, for example. Also a significant amount of work has been dedicated to running kaon production
targets at the AGS at $\stackrel{>}{\sim}$30 Tp/spill.

\section{AGS Kaon Experiments in the 1980's}
\label{sec_early}

 In the early 1980's two major kaon decay experiments took data. Looking 
for lepton family number violating decays,
E780 searched in the neutral kaon system for \klme\
and E777 searched in the charged kaon system for \kpme.
%Both experiments used decay in flight beams.

The experiment E780, which ran from 1985--88, followed earlier experiments by the same
group looking for T-violation in the transverse polarization of the $\mu^+$ in both
neutral~\cite{Akm3_0,Akm3_1} and charged~\cite{Akm3_2,Akm3_3} $K_{\mu3}$ 
decays (E696, E735) and $\epsilon'/\epsilon$~\cite{Aeps1} (E749) from 1978--84.
E780 set limits on the decays \klme, \klee\ and \klpee\  and 
observed the decay \klmm.~\cite{Amue1,Ap0ee1,Amue2} 
The experiment E777 set limits on the decay \kpme. It also set limits on \pme\ and \kpX\
and measured the branching ratio for \kpee~\cite{piX,pimue1,pimue2,piee1}.

In the mid-to-late 1980's a second round of experiments was started. 
Follow-up experiments to both E780 and E777 were commissioned, with modest modifications to the
apparatus and triggers, to focus on decays with  electrons in the final state.
E845 continued E780's work on \klpee~\cite{Ap0ee2,Ap0ee3} and
measured branching ratios for \kleeg~\cite{Aeeg}, \kleegg~\cite{Aeegg} and \kleeee~\cite{Aeeee}. 
This experiment discovered that \kleegg\ would be a significant background to \klpee.~\cite{Aeegg1}
The  experiment was 
optimized for many-body decays with electrons and photons. 
The spectrometer was shortened by moving the Cherenkov detector
inside the magnet and the size of the first drift chamber was expanded. New triggers 
were created. The kaon production angle
was changed from 0$^\circ$ to 2$^\circ$ to reduce rates and increase acceptance for tracks near the
neutral beam.  E845 took data during 1989. Results from E780 and E845 are summarized in 
Table~\ref{tab_780}.
\begin{table}[htb]
\begin{center}
\begin{tabular}{|l||l|c|}\hline
mode & Measurement & Comments \\ \hline\hline
\klme   & $<1.9\times 10^{-9}$ & (Ref~\citenum{Amue2}) \\ \hline
\klpee  & $< 5.5 \times 10^{-9}$ & (Ref~\citenum{Ap0ee3}) \\ \hline
\klee   & $<1.2\times 10^{-9}$ & (Ref~\citenum{Amue2}) \\ \hline
\kleeg  & $(9.1\pm0.4 ^{+0.6}_{-0.5}) \times 10^{-6}$ & 919 events (Ref~\citenum{Aeeg}) \\ \hline
\kleegg & $(6.6\pm3.2) \times 10^{-7}$ & 17 events (Ref~\citenum{Aeegg}) \\ \hline
\kleeee & $(3.07\pm1.25\pm0.26) \times 10^{-8}$ & 6 events (Ref~\citenum{Aeeee}) \\ \hline
\end{tabular}
\caption{Results from E780/E845.}\label{tab_780}
\end{center}
\end{table}

E851 improved the E777 
measurement of \kpee\ and measured the \pee\ branching ratio~\cite{pi0ee}.
The Cherenkov detector was optimized for electron detection on both sides of the spectrometer
and a new trigger was installed. E777 ran from 1986--88 and E851 ran in 1989. The results from
the E777 and E851 experiments are listed in Table~\ref{tab_777}.
\begin{table}[htb]
\begin{center}
\begin{tabular}{|l||l|c|}\hline
mode & Measurement & Comments \\ \hline\hline
\kpme  & $<2.1 \times 10^{-10}$                & (Ref~\citenum{pimue2}) \\ \hline
\pme   & $<1.6\times 10^{-8}$                  & (Ref~\citenum{pimue2}) \\ \hline
\pee   & $(6.9\pm2.3\pm0.6)\times 10^{-8}$     & 21 events (Ref~\citenum{pi0ee}) \\ \hline
\kpee  & $(2.75\pm0.23\pm0.13) \times 10^{-7}$ & 487 events (Ref~\citenum{piee1}) \\ \hline
\kpX   &                                       & \\
$\hookrightarrow X^\circ\rightarrow e^+ e^-$& $<4.5 \times 10^{-7}$  & $M_X<100$ MeV/c$^2$ (Ref~\citenum{piX}) \\ \hline
\kpX   &                                       & \\
$\hookrightarrow X^\circ\rightarrow e^+ e^-$& $<1.5 \times 10^{-8}$   & $150<M_X<340$ (Ref~\citenum{piee1}) \\ \hline
\end{tabular}
\caption{Results from E777/E851.}\label{tab_777}
\end{center}
\end{table}

Two other `2$^{nd}$ generation' experiments also started taking data in the late 1980's: E791 and E787. 
The primary 
purpose of E791 was a further improvement in sensitivity in the search for \klme. This experiment had
a two arm spectrometer, with high precision momentum measurements and good $e$ and $\mu$
identification. The goal of the experiment was a sensitivity to this lepton flavor violating decay of $10^{-12}$.
E791 had an engineering run in 1988 and long physics runs in 1989--90. The physics results
from E791 are listed in Table~\ref{tab_791}.~\cite{mue0,mue1,mumu1,mumu2,mue2,ee1,mumu3} 
\begin{table}[htb]
\begin{center}
\begin{tabular}{|l||l|c|}\hline
mode & Measurement & Comments \\ \hline\hline
\klme   & $<3.3\times 10^{-11}$ & (Ref~\citenum{mue2}) \\ \hline
\klee   & $<4.1 \times 10^{-11}$ & (Ref~\citenum{ee1}) \\ \hline
\klmm   & $(6.86\pm0.37)\times 10^{-9}$ & 707 events (Ref~\citenum{mumu3}) \\ \hline
\end{tabular}
\caption{Results from E791.}\label{tab_791}
\end{center}
\end{table}

Another new experiment,
E787, was designed to search for the SM allowed decay \kpnn. 
This experiment needs to identify the single, isolated $\pi^+$ track from \kpnn. 
The experiment used a stopped K$^+$ beam,
rather than the decay in flight beams used by the other experiments. It has a $4\pi$
hermetic photon veto system, good kinematic resolution (range, momentum and energy) and $\pi^+$ identification
through the decay chain $\pi^+\rightarrow\mu^+\rightarrow e^+$. 
The first phase of E787~\cite{det} had an engineering run in 1988 and took physics data from 1989--91. 
The results from the first phase of
E787 are given in Table~\ref{tab_787}.~\cite{higgs,pnn1,pgg1,pi0_nunu,pi0_gX,pnn2,pnn3,pnn4,pnn5,pgg2,pmm}
The upgraded E787 experiment and detector will be discussed in more detail in section~\ref{sec_e787}.
\begin{table}[htb]
\begin{center}
\begin{tabular}{|l||l|c|}\hline
mode & Measurement & Comments \\ \hline\hline
\kpnn   & $<2.4\times10^{-9}$ & (Ref~\citenum{pnn5}) \\ \hline
\kpX   & $<5.2\times10^{-10}$ & (Ref~\citenum{pnn5}) \\ \hline
\kpmm   & $(5.0\pm0.4\pm0.7\pm0.6)\times10^{-8}$ & 207 events (Ref~\citenum{pmm}) \\ \hline
\kpgg   & $(1.1\pm0.3\pm0.1)\times10^{-6}$ & 31 events (Ref~\citenum{pgg2}) \\ \hline
\pgX   & $<5\times 10^{-4}$ & (Ref~\citenum{pi0_gX}) \\ \hline
\pnn & $<8.3 \times 10^{-7}$ & (Ref~\citenum{pi0_nunu}) \\ \hline
\kmnmm   & $<4.1\times10^{-7}$ & (Ref~\citenum{higgs}) \\ \hline
\end{tabular}
\caption{Results from the first phase of E787.}\label{tab_787}
\end{center}
\end{table}
Two new results from the first 
phase of the experiment have recently been published in PRL: \kpgg~\cite{pgg2} and    
\kpmm~\cite{pmm}.

\begin{figure}[hbt]
\center\epsfig{file=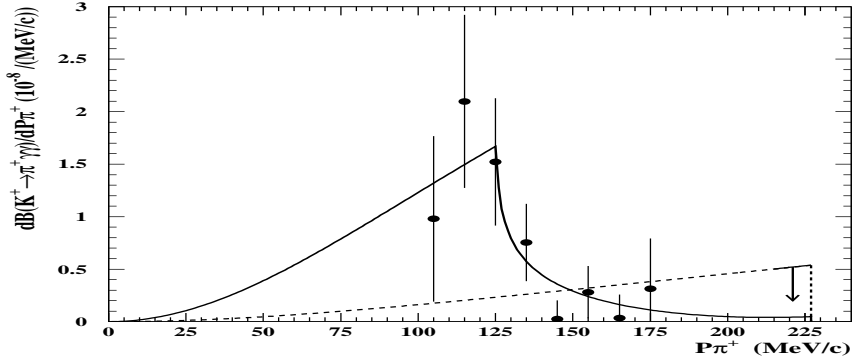,width=5.in,height=2.25in,angle=0}
\caption{\kpgg: 31 events. The solid line is the best \XPT\ fit. The dashed line is a phase-space distribution
normalized to the 90\%CL limit in the $\pi\gamma\gamma$1 region.}\label{f787_pgg}
\end{figure}

E787 reported the first observation of the decay \kpgg.~\cite{pgg2}
The branching ratio and invariant mass $M_{\gamma\gamma}$ for the decay \kpgg\ are predicted by
\XPT.~\cite{eck88,eck90,cheng,bruno,eck93}
 The prediction for $M_{\gamma\gamma}$ is striking. Instead of a smooth phase space distribution, there
is a sharp turn-on above $M_{\gamma\gamma}$ = 2$M_\pi$. The measured spectrum verifies this
prediction, as can be seen from the momentum spectrum of the $\pi^+$ shown in figure~\ref{f787_pgg}. 
The \kpgg\ trigger had two components: one 
accepting events with a $P_\pi$ above the K$_{\pi2}$ peak ($\pi\gamma\gamma$1) and one accepting
$P_\pi$ below the K$_{\pi2}$ peak ($\pi\gamma\gamma$2). There were 31 events observed in the $\pi\gamma\gamma$2
sample and none observed in the $\pi\gamma\gamma$1 sample.
The model independent branching ratio is 
\begin{equation}
B(\kpgg: 100<P_{\pi^+}<180 {\rm MeV/c}) = (6.0\pm1.5\pm0.7)\times10^{-7} .
\end{equation}
The fit to the branching ratio and spectrum favors the \XPT\ parameter of \^{c} = $1.8\pm0.6$ with
the so called `unitarity corrections'~\cite{capp,cohen}. The total branching ratio (assuming \^{c}=1.8) is 
\begin{equation}
B(\kpgg) = (1.1\pm0.3\pm0.1)\times10^{-6} .
\end{equation}
In the $\pi\gamma\gamma$1 region a 90\%CL limit of 
\begin{equation}
B(\kpgg:P_{\pi^+}>215 {\rm MeV/c}) < 6.1\times10^{-8} 
\end{equation}
was set.

\begin{figure}[htb]
\begin{minipage}[htb]{.44\linewidth}
\centering\epsfig{file=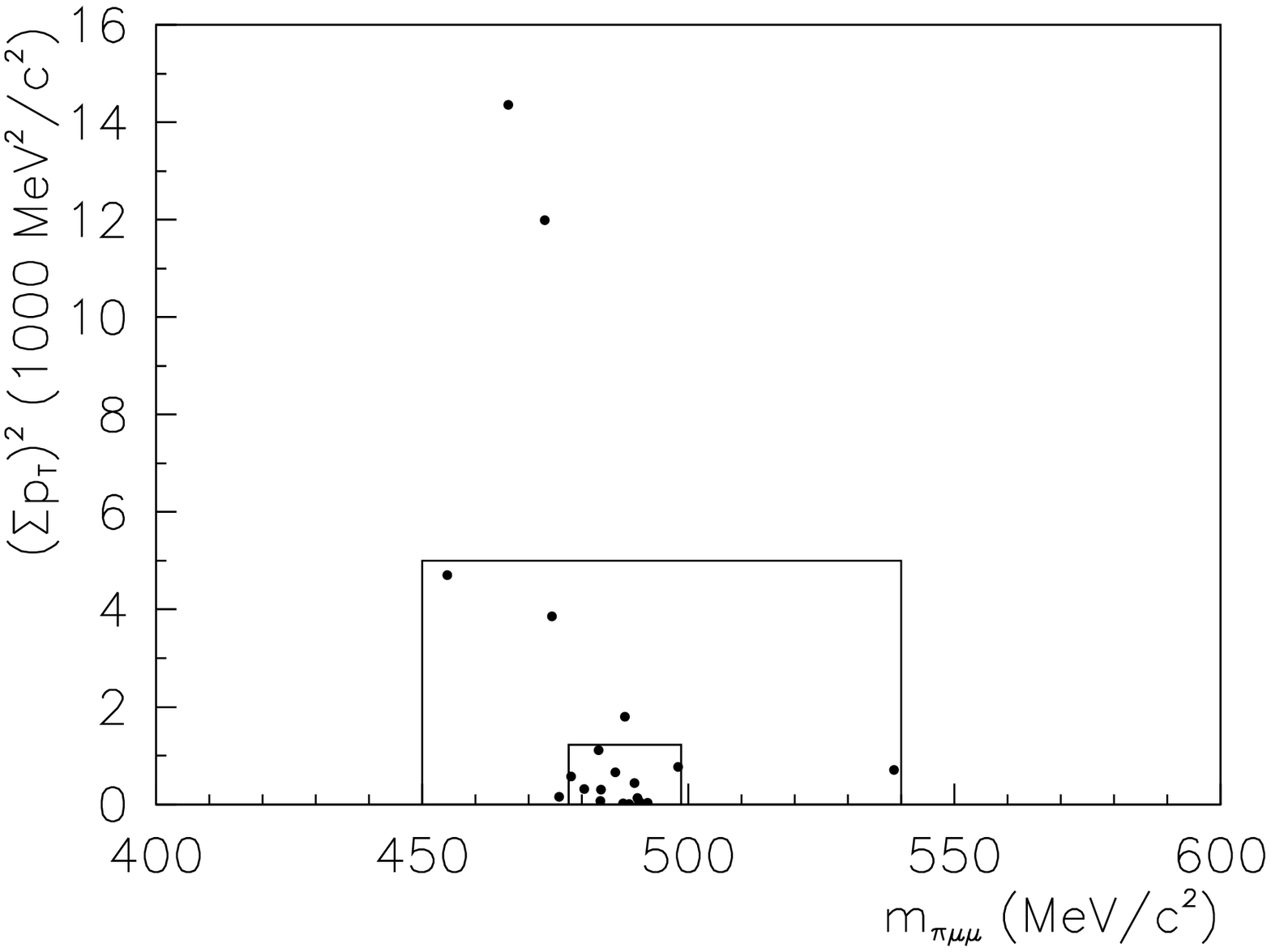,width=\linewidth,angle=0}
\end{minipage}\hfil 
\begin{minipage}[htb]{.44\linewidth}
\centering\epsfig{file=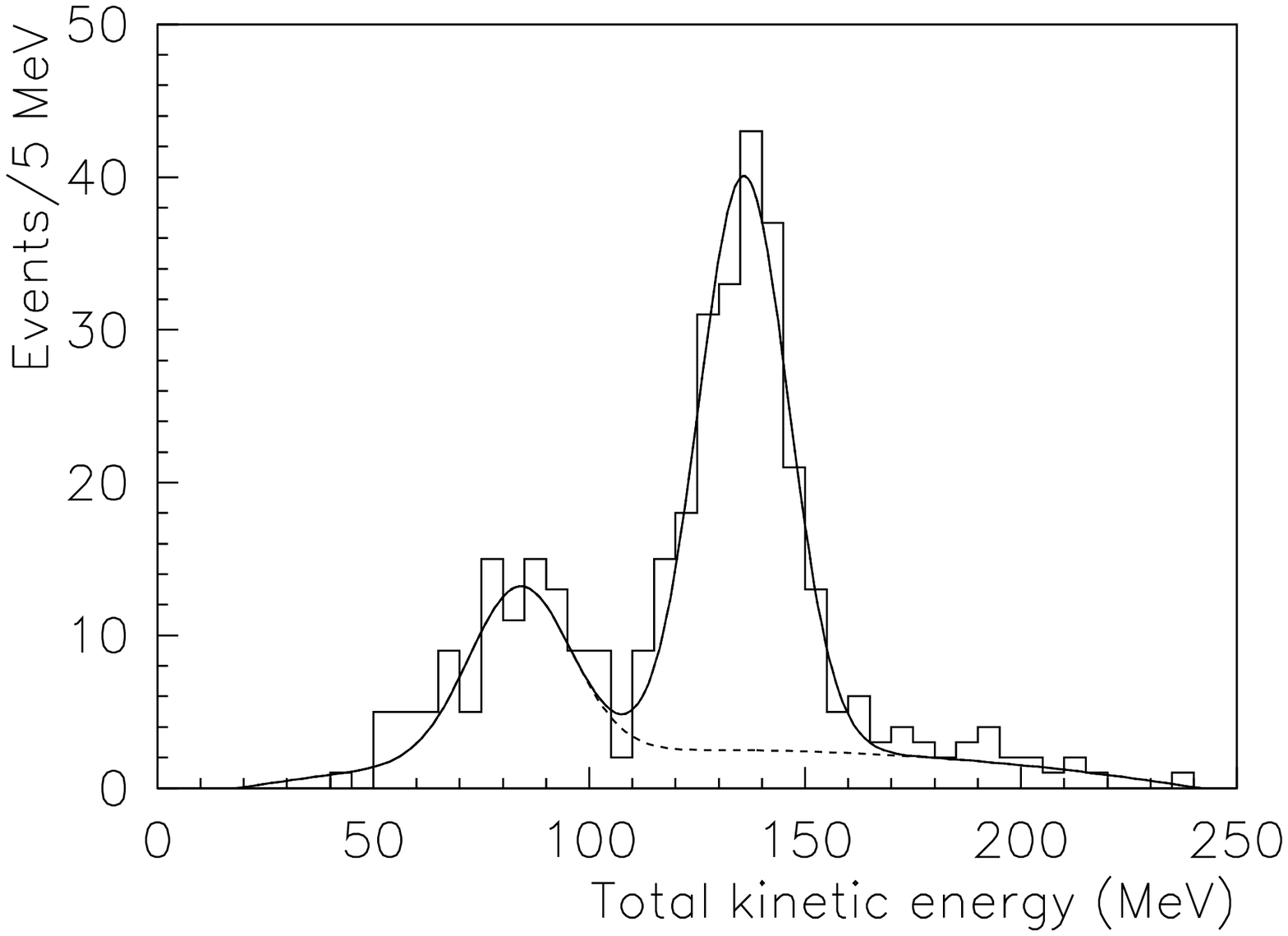,width=\linewidth,angle=0}
\end{minipage}
{\caption{\label{pmm1a}\label{pmm1b}Final event candidates for \kpmm: a) 3-track events and b) 2-track events.}}
\end{figure}

E787 has also reported the first observation of the decay \kpmm.~\cite{pmm}
Two separate analyses of this decay were made. In the first all three tracks were fully reconstructed.
In the second the minimum requirement for the third track was that its energy was measured.
In the first analysis a total of 10.6$\pm$4.7 events were observed and in the second
analysis 196.0$\pm$16.7 events were observed (see figure~\ref{pmm1b}). 
These two analyses were combined to give a  branching ratio for this decay of:
\begin{equation}
B(\kpmm) = (5.0\pm0.4^{stat}\pm0.7^{syst}\pm0.6^{theor})\times10^{-8} . 
\end{equation}
This branching ratio implies a value for the \XPT\ parameter~\cite{eck87}
$w_+ = 1.07 \pm 0.07$. This value is larger than that derived from \kpee: $w_+ = 0.89^{+0.24}_{-0.14}$ (Ref~\citenum{piee1}).

\section{Current Kaon Experiments}

There are three major rare kaon decay experiments running or analyzing data currently.
All three experiments obtained sizable physics data sets from the 1995 and 1996 runs. Due to
the short length of the 1997 run (funding for only 8 weeks), E871 decided to concentrate their efforts on analysis of
existing data and, assuming they do not find any \klme\ events, will probably not take any more data.
The E865 experiment chose to use the short 1997 run to collect some special data that is not
compatible with their \kpme\ running. The E787 experiment, after some additional efficiency improvements,
was able to collect additional \kpnn\ data ($\sim$60\% of the sensitivity of the 1996 run).
Both E865 and E787 are planning to collect a significant data sample from the 1998--99 running period.

\subsection{E787}
\label{sec_e787}

\begin{figure}[htb]
\center\epsfig{file=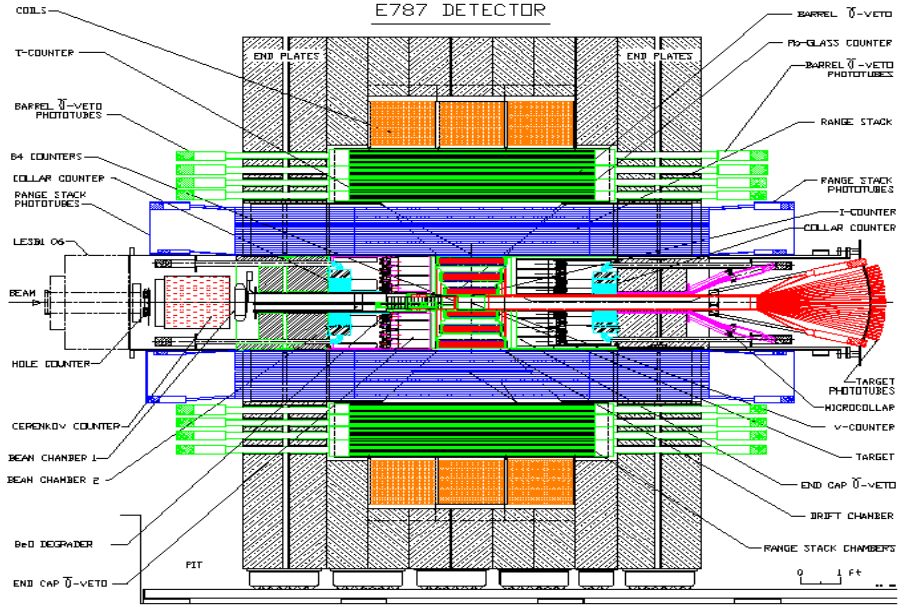,width=5.5in,height=3.5in,angle=90}
\caption{E787 experimental layout.}\label{f787_det}
\end{figure}

The E787 experiment was first proposed in 1983 and took significant physics data from 1989--91.
The experiment underwent a major upgrade from 1992--94 (keeping the number E787). 
A drawing of the new detector is shown in figure~\ref{f787_det}.
The detector is located in the C4 beam line at the AGS. The C4 beam line, or Low Energy Separated Beam (LESB3),~\cite{lesb3}
transmits kaons of up to 830 MeV/c. At 800 MeV/c it can transmit up to $6\times10^6 K^+$ with
a $K/\pi$ ratio of 3:1 for 10$^{13}$ protons. The $K^+$ are stopped in a scintillating fiber target in the center of the detector.
The detector itself covers close to $4\pi$ solid angle and is in a 10 kG magnetic field.

The decay \kpnn\ is interesting because of its sensitivity to the CKM matrix element \vtd,
and because the theoretical uncertainty in calculating the branching ratio is very 
small~\cite{hagelin,bb94,marciano,bb93,rein,lu,geng,faijfer}: 
only 7\%~\cite{bf}. From current values of SM parameters the branching ratio is expected to be
B(\kpnn) = (0.6--1.5)$\times10^{-10}$ (Ref~\citenum{bbl}).

The observation of \kpnn\ is complicated by two significant difficulties: 1) the small branching
ratio $\sim10^{-10}$ and 2) the poor experimental signature.
Since there are no major $K^+$ decay modes with a $\pi^+$ with momentum above the $K_{\pi2}$
momentum (205 MeV/c), this experiment chooses to look for \kpnn\ with a high momentum $\pi^+$.
Excellent resolution on $P_\pi^{cm}$ is obtained by working in the $K^+$ center of mass, 
so E787 uses a stopped $K^+$ beam.
Since the signature of \kpnn\ is poor, all backgrounds need to be well identified. 
Most $K^+$ decays with a $\pi^+$ also have a $\pi^\circ$, including the $K_{\pi2}$ decay (BR=21\%).
These are identified and vetoed with a fully active $4\pi$ detector, where any activity in time with the
$\pi^+$ track, not associated with the track, is vetoed. The other major background comes from the
$K_{\mu2}$ decay (BR=64\%), which is rejected for not having a $\pi^+\rightarrow\mu^+\rightarrow e^+$ decay chain.
Both of these decays are mono-energetic and can also be rejected based on kinematics (range, momentum 
and energy).  The other significant background comes from either a $\pi^+$ in the beam scattering into the
detector or from charge exchange. The scattered pions are rejected by good particle identification in the beam,
including reconstruction of both the kaon and the pion in the target, and by requiring that
the `kaon' decay at rest in the target, (i.e. a `delayed coincidence' requiring that the $\pi^+$ track occur later than the
incident $K^+$ track). The charge exchange background, $K^+ + n \rightarrow p + K_L$ followed by $K_L \rightarrow \pi^+ \ell^- 
\nu_\ell$, is rejected by identification of the extra particles (the proton and lepton)
and by the delayed coincidence. Any $K_L$ decay that satisfies the delayed coincidence,
yet remains in the target, has to come from a very low momentum $K_L$.

The beam line was upgraded before the 1992 run (LESB1 $\rightarrow$ LESB3), and
the detector was upgraded between 1992--94:
\begin{itemize}
 \item The incident kaon flux was increased and the K/$\pi$ ratio was increased from 1:3 $\longrightarrow$ 3:1.~\cite{lesb3}
 \item The drift chamber was rebuilt. The new Ultra Thin Chamber~\cite{UTC} (UTC) has $\times$2
better momentum resolution $\sim$1\%  and $\times$2 better $z$-resolution. 
The $z$ is measured from the
charge distribution on helical cathode strips. The total amount of material in the active momentum
measuring section of chamber is $2\times10^{-3}X_\circ$.
 \item The Pb-scintillator endcaps (EC) were replaced. The new undoped CsI endcaps~\cite{CsI1,CsI2,CsI3} are fully active (less
sensitive to inefficiencies due to photonuclear interactions) and have substantially more light
output. The timing resolution is $\times$2--3 better than with the old endcaps. This is critical
in the EC, which has the highest rates in the detector.
 \item Transient digitizers made from GAs CCD's were installed to provide improved timing and multiple
pulse finding capability (particularly to find pions on the tails of kaons in the target). The CCD's~\cite{CCD}
have 8 bit resolution with 500 MHz sampling. These have been used to instrument the target, endcaps and beam
elements.
 \item Very low mass Range Stack Straw Chambers (RSSC) were installed. These have better $z$-resolution than
the previous proportional counters, and even more important have significantly less dead material.
 \item The range stack (RS) scintillators were demultiplexed. Previously several counters were multiplexed
into one PMT. Now each counter (21 layers $\times$ 24 sectors) is read out on both ends by PMT's. This provides
more light and finer resolution for dE/dx measurements.
 \item A new target with $\times$4 more light and significantly less dead material was built.
 \item Weak regions in the photon veto system were filled by moving the EC's closer to the target and by adding
new systems at smaller angles (the collar, micro-collar and lead-glass (PbG) detectors).
 \item A new trigger/DAQ system was built.~\cite{e787_daq1,e787_daq2,e787_daq3} 
%The new trigger system resides in Fastbus modules.
The DAQ is Fastbus based, with frontend readout into Slac Scanner Processors (SSP). The data is transferred via the
Cable Segment to a Master SSP, where the event building is done. The data is transferred from the Master SSP to
an SGI Challenge computer via a Branch Bus to Fastbus Converter.~\cite{e787_daq4} The data transfer rate has been demonstrated
to be $\stackrel{>}{\sim}$24 Mbytes per spill 
(Data transfer to the SGI occurs in the 2 seconds between spills. Spills occur every 3.6 sec.)
The deadtime, which is dominated by the readout of the front end modules into the SSP's, is $\sim$20\%.
 \item New beam elements were built to further aid identification of scattered beam pions. A new Cherenkov
counter has larger acceptance for all beam particles. A second beam chamber was added and a PbG counter was
installed for additional photon veto coverage in the beam direction and for beam pion identification.
\end{itemize}

\begin{figure}[htb]
\center\epsfig{file=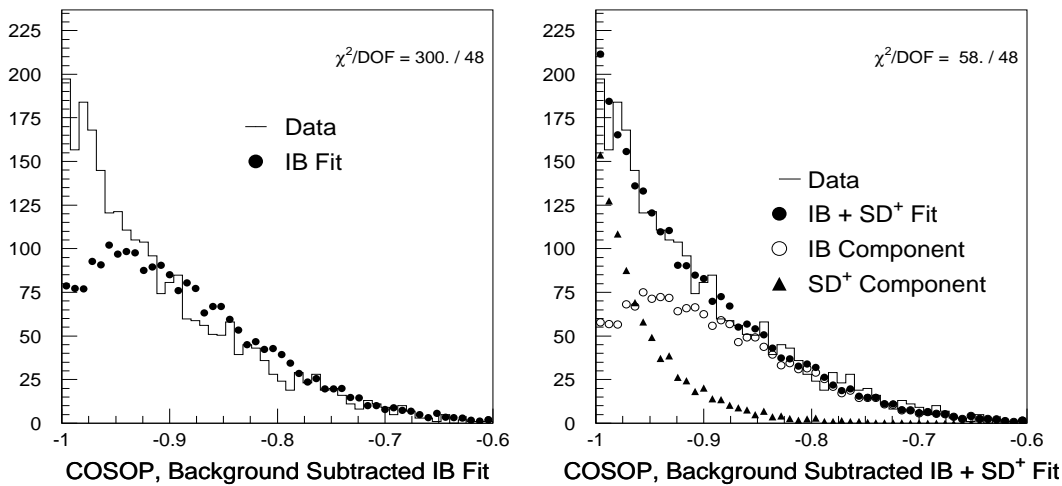,height=2.5in,width=5.5in,angle=0}
\caption{Distribution of $\cos \theta_{\mu \gamma}$ for E787 \kmng\ candidates. A fit to IB alone is
shown in (a), and a fit to both IB and \sdp\ is shown in (b).}\label{f787_kmng}
\end{figure}

An engineering run in 1994 produced the first physics result: the first observation of
the Structure Dependent (\sdp) component of the radiative decay \kmng~\cite{kmug1,kmug2}.
The \kmng\ decay is dominated by Inner Bremsstrahlung (IB), which has been well measured;~\cite{akiba}
however, the structure dependent component had not previously been seen.
Two days of data taking with a special trigger, selecting a high energy muon accompanied by
a high energy photon, yielded a total of $\sim$2700 $K_{\mu2\gamma}$ events with a background of 100 
events ($K_{\mu3}$ and $K_{\pi2}$ with
a missing photon and $K_{\mu2}$ with an accidental photon). Of the 2700 $K_{\mu2\gamma}$ events a clear
signal of about 900 events from the \sdp\ component is seen in figure~\ref{f787_kmng}. 
From a more sophisticated analysis, fitting E$_{\mu^+}$ and E$_{\gamma}$ to  IB, SD$^\pm$ and the interference terms, 
the branching ratio for the \sdp\ decay is: 
\begin{equation}
B(\sdp)  =  (1.33\pm0.12\pm0.18)\times10^{-5} 
\end{equation}
and the form factor $|F_V+F_A| = 0.165\pm0.007\pm0.011$.
This can be compared to the $O(p^4)$ \XPT\ calculation of $F_V+F_A = -0.137\pm0.006$ and 
B(\sdp)=9.22$\times10^{-6}$ (Ref~\citenum{bijens}). An $O(p^6)$ calculation is underway.~\cite{bij2,ametller}

\begin{figure}[htb]
\begin{minipage}[htb]{.44\linewidth}
\centering\epsfig{file=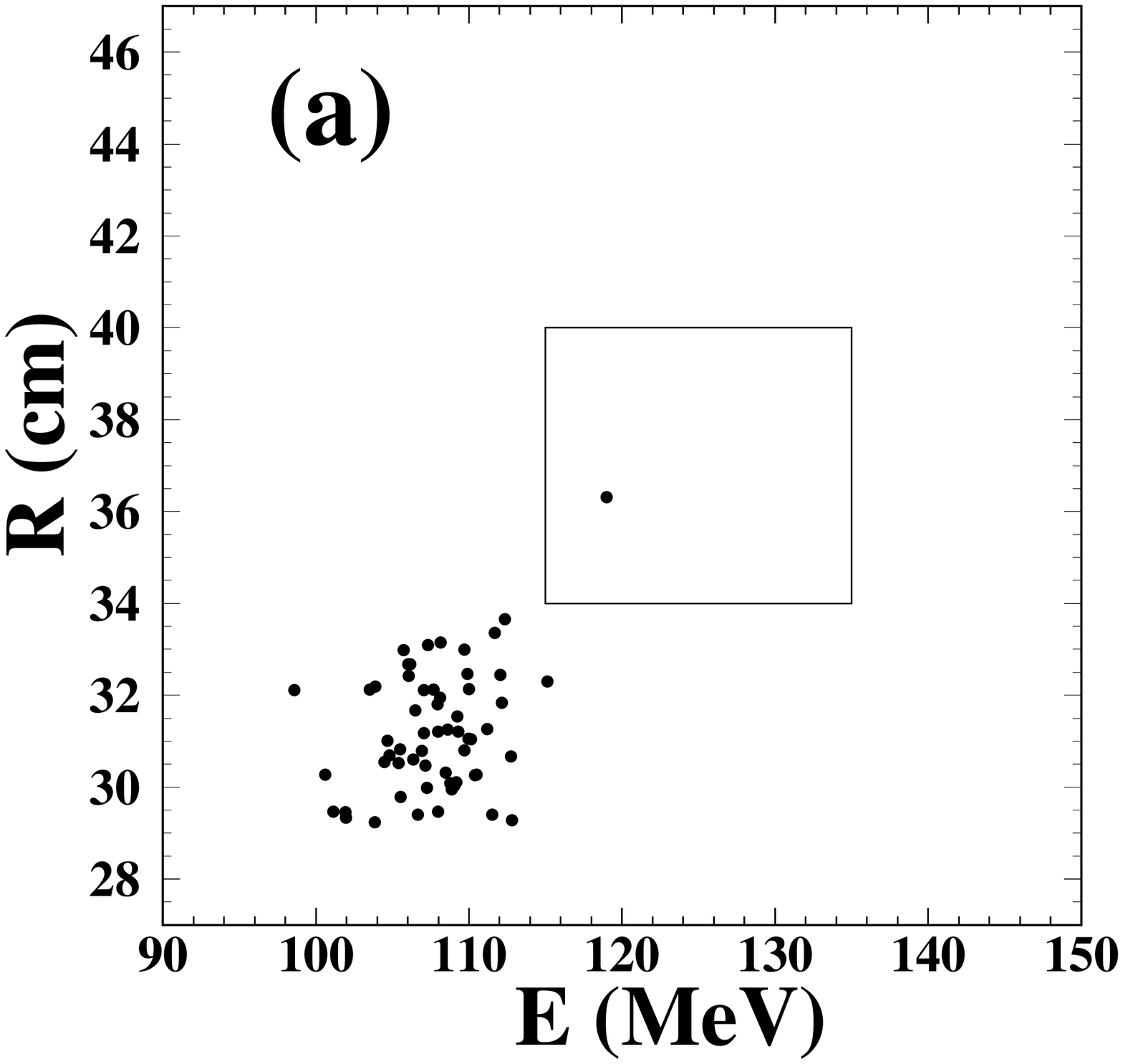,width=\linewidth,height=2.5in,angle=0}
\end{minipage}\hfil 
\begin{minipage}[htb]{.44\linewidth}
\centering\epsfig{file=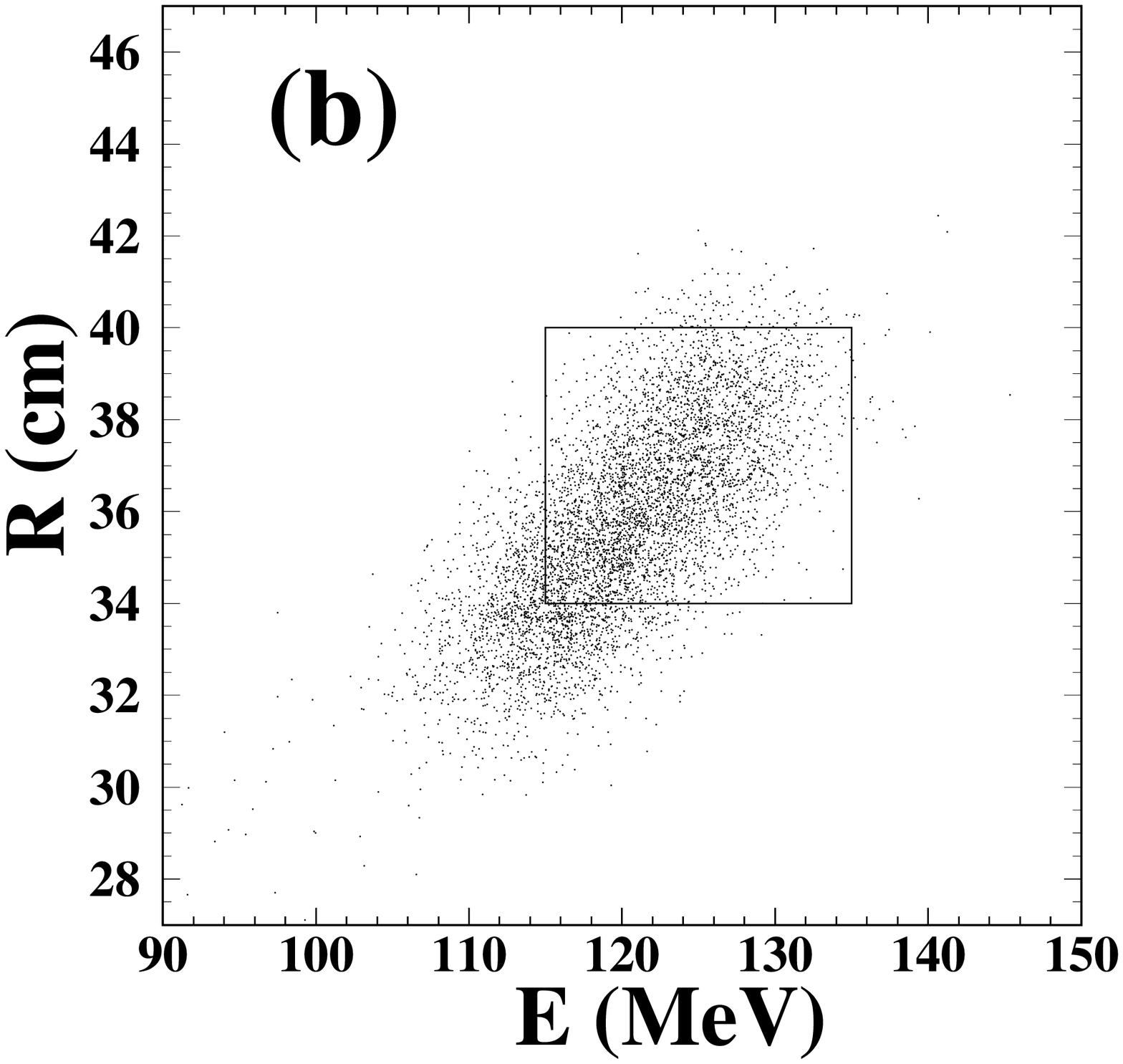,width=\linewidth,height=2.5in,angle=0}
\end{minipage}
{\caption{\label{787_pnn1}Final event candidate for \kpnn: a) Data b) Monte-Carlo} }
\end{figure}

\begin{figure}[htb]
\center\epsfig{file=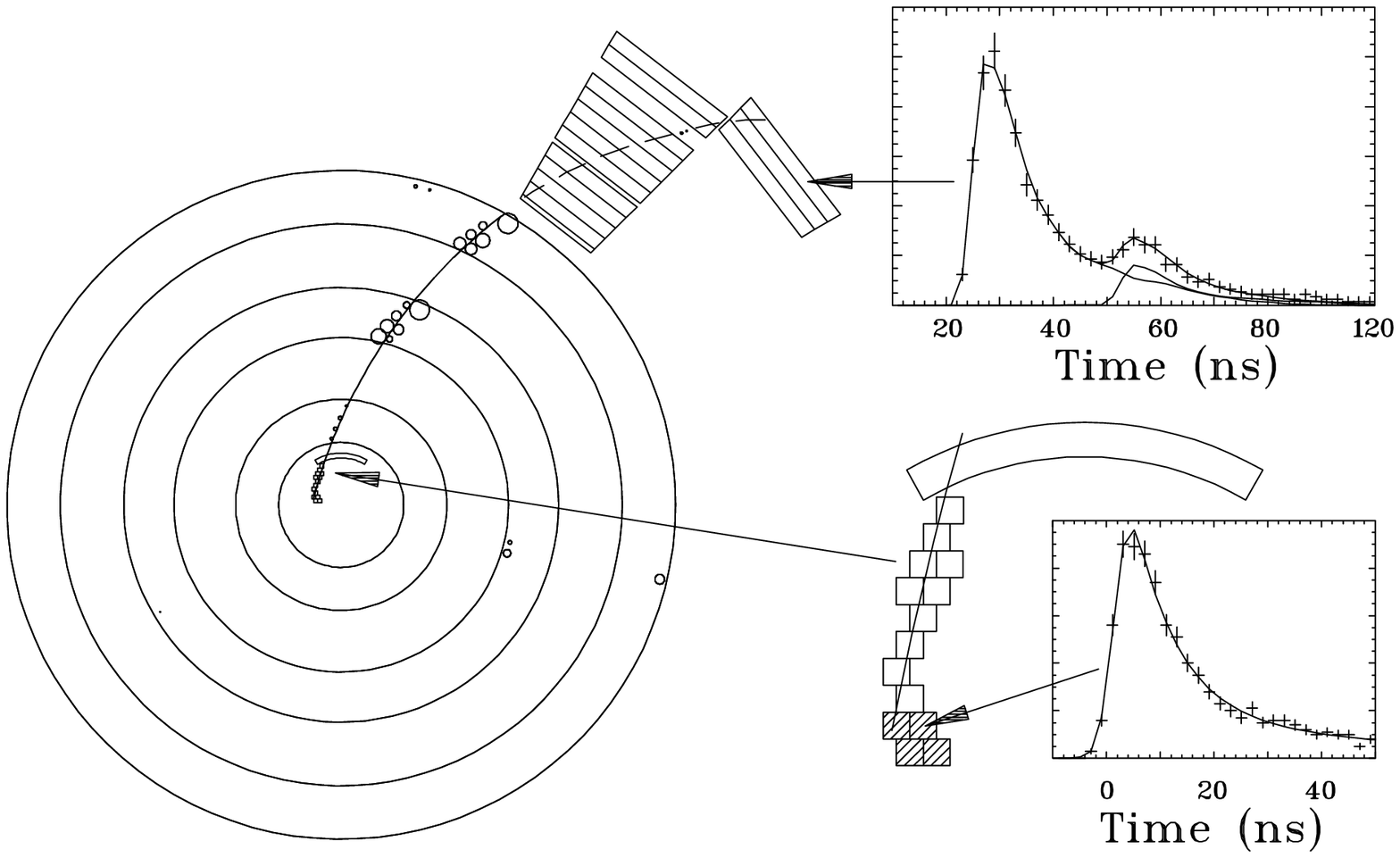,width=4.0in,height=2.5in,angle=0}
\caption{\kpnn\ event.}\label{787_pnn2}
\end{figure}

The first results from the search for \kpnn\ with the upgraded detector have recently been
published for the 1995 data set.~\cite{pnn6} The range and energy of event candidates passing 
all other cuts is shown in figure~\ref{787_pnn1}.
One event consistent with the decay \kpnn\ was observed. The expected background from all sources is $0.08\pm0.03$ events.
This event is in the so called `Golden Region' where the expected background is $0.008\pm0.005$ and with 55\% of
 the acceptance of the signal region. A reconstruction of the event is shown in figure~\ref{787_pnn2}.
%The event has no significant energy deposits in time with the $\pi^+$. 
There is a clean $\pi^+\rightarrow\mu^+$ decay at 27.0 ns, as can be seen in the upper insert in
figure~\ref{787_pnn2}; there is also a clean $\mu^+\rightarrow e^+$ decay at 3201.1 ns. The $K^+$ decay occurred at 23.9 ns.
There is no significant activity anywhere else in the detector at the time of the $K^+$ decay.
The branching ratio for \kpnn\ implied by the observation of this event is
B(\kpnn) = 4.2$^{+9.7}_{-3.5} \times 10^{-10}$,
compared to the expected value from the SM calculation of (0.6--1.5)$\times10^{-10}$.
The expected sensitivity from the data already on tape (1995--97) is $\sim$2.4 times that
of the 1995 data alone. Additional improvements from more efficient analysis software are also
expected. Results of the analysis of the larger data set are expected within the next year.
With improved running conditions and a long running period in 1998--99, the E787
sensitivity for \kpnn\ should extend well below the SM level.

\subsection{E865}

The E865 experiment has been substantially upgraded from the previous experiments (E777/E851).
The primary goal is a search for the `forbidden' lepton flavor violating  decay \kpme. 
The proposed sensitivity for \kpme\ is $10^{-12}$ or $\sim\times$70 greater than for E777. The
experiment is running in the A2 beam line with an unseparated 6 GeV/c K$^+$ beam  containing
$5\times10^7$  K$^+$/spill and $1\times10^9$ $\pi^+$ and $p^+$ per spill. Approximately 9\%
of the $K^+$ decay in the 5 m decay volume. The detector rates
are $\sim$50 M/spill and the first level trigger rate (3 charged tracks) is 2M/spill. After higher level
triggers requiring a good $e^-$ and $\mu^+$ $\sim$1000 events are written to tape each spill.

\begin{figure}[htb]
\center\epsfig{file=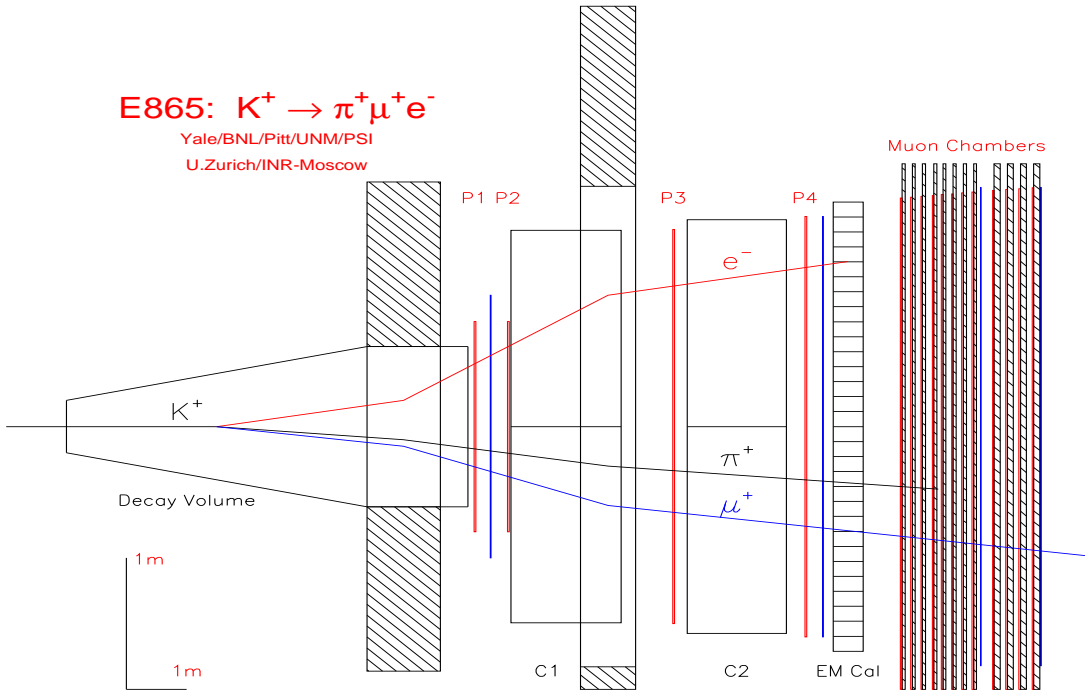,width=5.5in,height=3.5in,angle=-90}
\caption{E865 experimental layout.}\label{f865_exp}
\end{figure}

A drawing of the
experimental apparatus is shown in figure~\ref{f865_exp}. 
Good particle identification is essential. The first magnet separates $K^+$ decay products by charge,
so that the left side of the spectrometer is optimized for small 
$e^-$ mis-identification probability and the right
side is optimized for small $\mu^+$ and $\pi^+$ mis-ID probability. 
A coincidence between the two hydrogen Cherenkov counters on the left side and the Pb-scintillator
calorimeter (Shashlyk design: Pb and scintillator plates readout by a wavelength shifting fiber) 
is required to provide small probability of mis-identifying a particle as an electron.
The Cherenkov counters
on the right side are filled with a lower threshold gas (CO$_2$ in 1995 and CH$_4$ in 1996) to
more efficiently veto $e^+$.
The trigger rate is reduced significantly by
requiring an $e^-$ on the left side.
Background from $K_{\pi2}$ and $K_{\mu3}$ decays with a $\pi^{\circ}$ Dalitz decay are suppressed by
assuring that $e^+$ on the right side are never mis-identified as $\mu^+$'s or $\pi^+$'s.
The spectrometer consists of 4 sets of Multi-Wire proportional Chambers (P1--4) surrounding the
spectrometer magnet.
The reconstructed mass resolution for $K_{\pi3}$ is $\sigma_{M_{K}}=2.3$ MeV/c$^2$. 
%The measured $e^-$ mis-ID is XXX$\times10^{-7}$ and the $\pi^+$ mis-ID is XXX$\times10^{-6}$.

A number of upgrades to the E777 experiment include:
\begin{itemize}
  \item A substantially improved  beamline compared to the one used for E777. With $\times$5 more
$K^+$, the detector rates are held constant due to improved optics and collimation.
  \item The spectrometer acceptance is increased by $\times$3 by increasing the detector sizes.
  \item A 2$^{nd}$  muon arm was added for increased acceptance. 
  \item An additional Y-view was installed in each chamber for more efficient online and offline event reconstruction.
  \item Graphite coated Mylar HV foils were used to reduce multiple scattering.
\end{itemize}

The expected sensitivity for \kpme\ from the 1995--96 runs is $\sim\times$5 greater than E777.
For \kpee\ and \pee\ the sensitivity is 
$\sim\times$10 greater ($\sim7000$ \kpee events). 
In addition, E865 is sensitive to the decay \kpmm\ and to the radiative decays
\kmnee\ and \kenee.

\begin{figure}[htb]
\center\epsfig{file=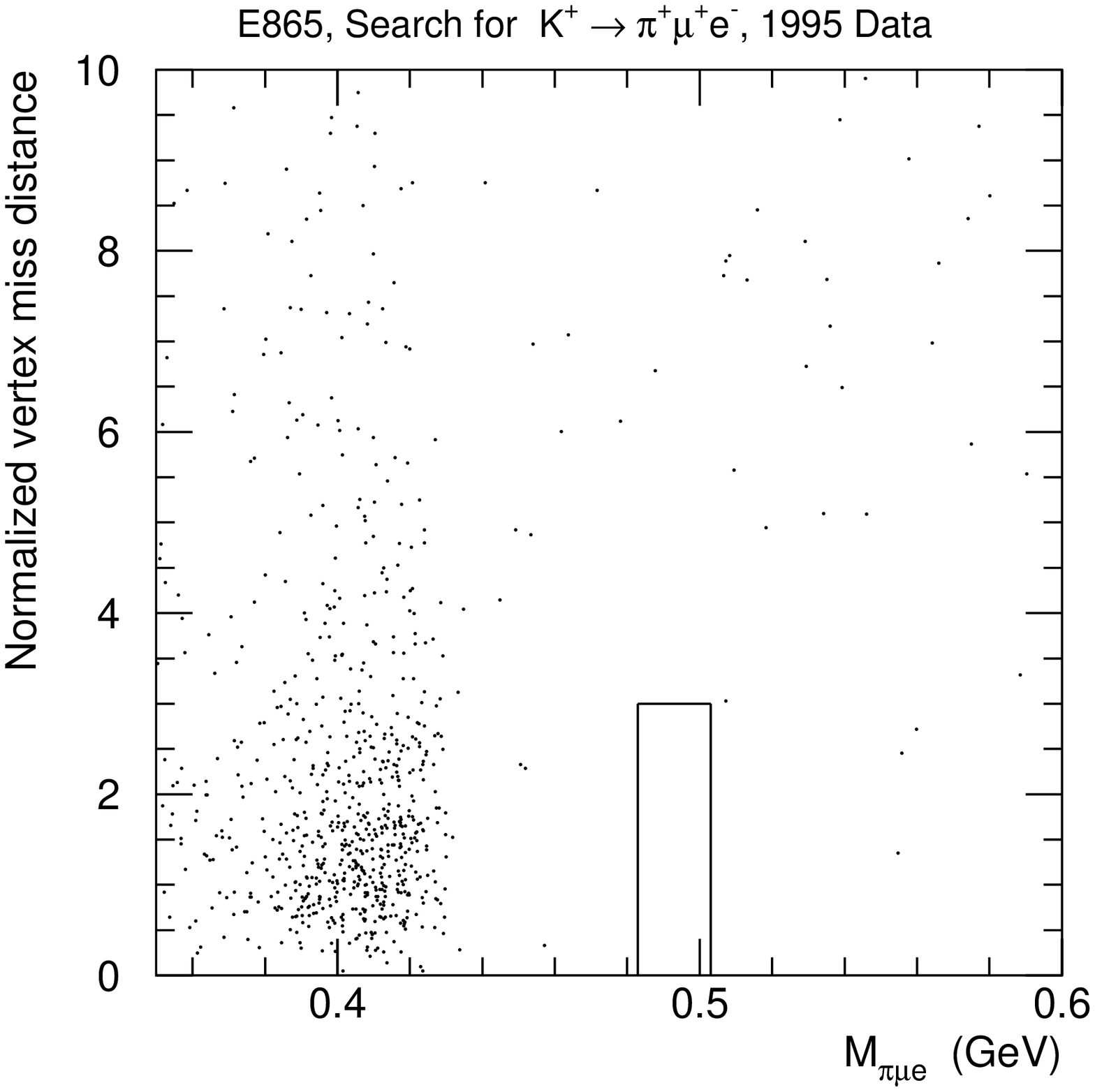,width=5.0in,height=2.in,angle=0}
\caption{E865 preliminary \kpme\ data from 1995.}\label{f865_pme}
\end{figure}

For the primary mode, the lepton flavor violating  \kpme\ decay, a preliminary analysis of the 1995 data 
shows no \kpme\ events (see figure~\ref{f865_pme}).
The plot shows the reconstructed mass $M_{\pi\mu e}$ vs the separation of the three tracks at the vertex. No events are 
seen in the box at $M_K$ and small vertex separation.
E865 has a small data sample for \kpme\ from 1995 and a much larger sample from 1996 ($\sim\times$4). 
The expected single event sensitivity from the combined 1995--96 data set is $\sim1.6\times10^{-11}$
The  data set should be increased  by another factor of 3 in the 1998--99 run.

\begin{figure}[htb]
\center\epsfig{file=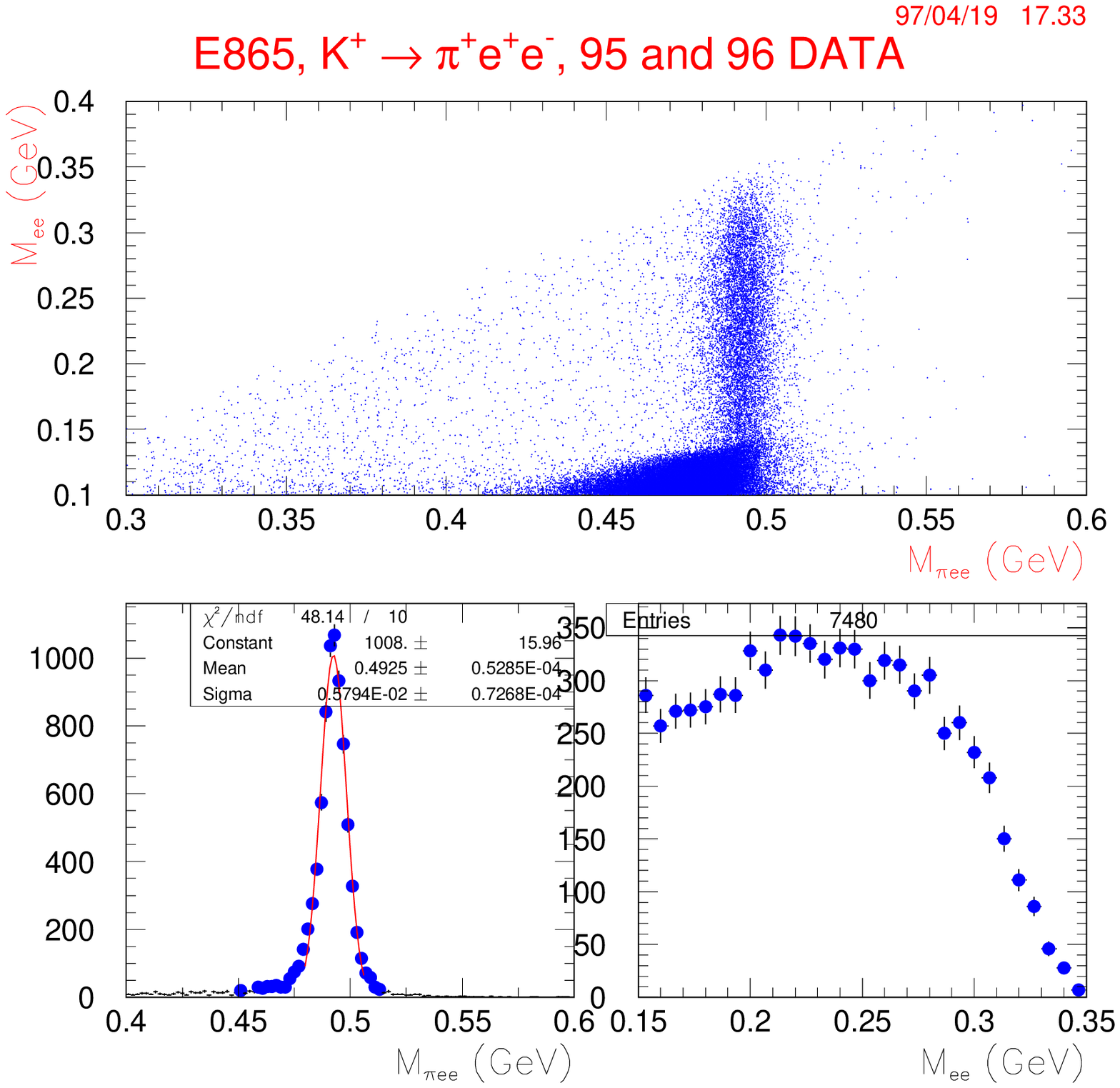,width=5.0in,height=2.75in,angle=0}
\caption{E865 preliminary \kpee\ data from 1995--96.}\label{f865_pee}
\end{figure}

The decays \kpee\ and \kpmm\ are interesting as tests of \XPT. Both the branching ratios and $M_{\ell\ell}$
for these two decays are expressed in terms of a single parameter $w_+$, as discussed previously in section~\ref{sec_early}. 
A preliminary analysis of the
combined 1995--96 data set for the \kpee\ decay is shown in figure~\ref{f865_pee}.
The first plot in figure~\ref{f865_pee} shows the reconstructed `kaon' mass, $M_{\pi ee}$ vs. the di-electron mass $M_{ee}$.
The \kpee\ signal can be seen at $M_{\pi ee} = M_K = 494$ MeV/c$^2$ and $M_{ee}>M_{\pi^\circ}$.
The remaining two plots show the reconstructed kaon mass and the di-electron mass for the \kpee\ signal.
There are more than 7000 events, as compared to the previous E777 sample of $\sim$500 events.~\cite{pi0ee}

\begin{figure}[htb]
\center\epsfig{file=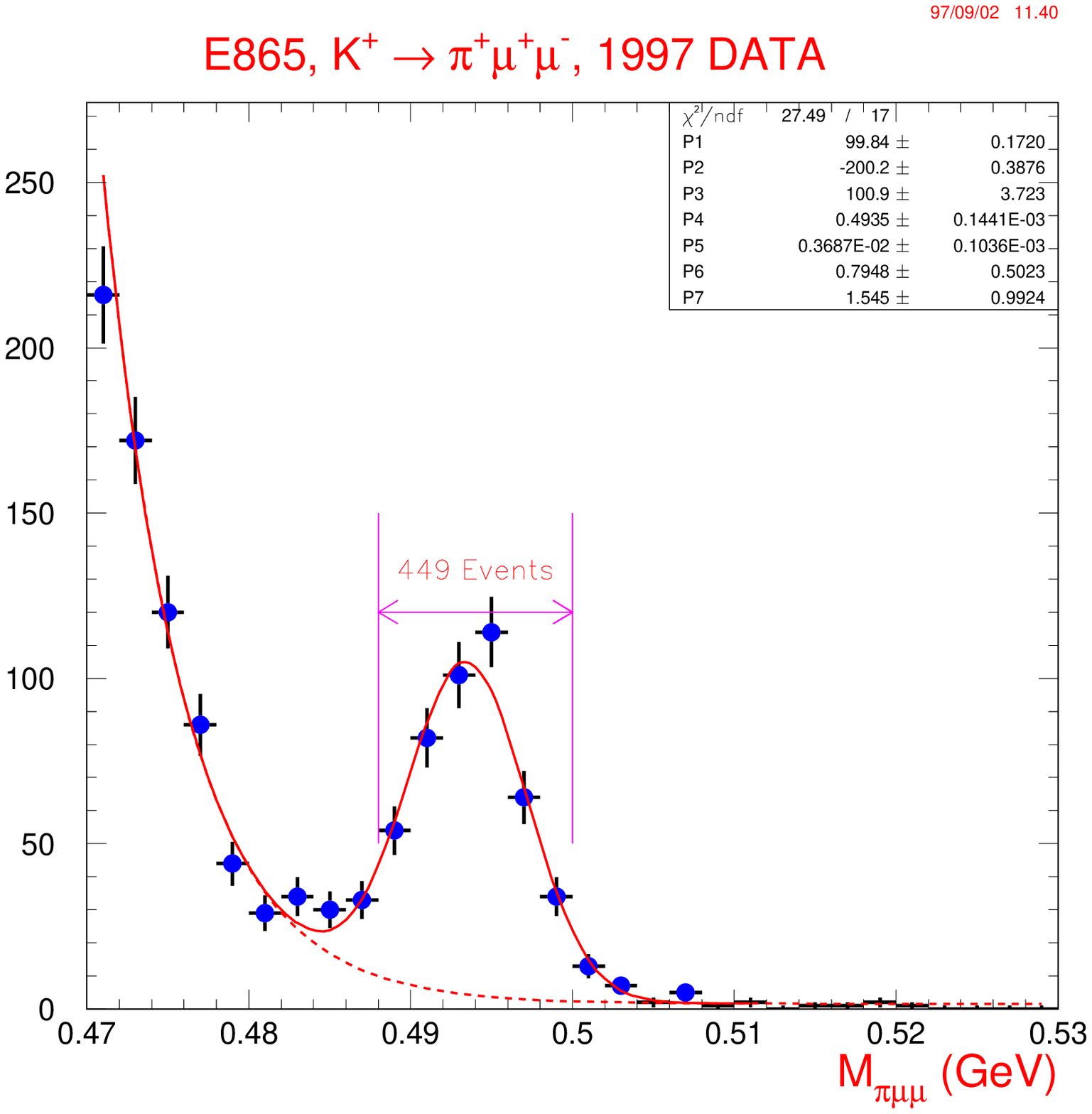,width=4.0in,height=3.0in,angle=0}
\caption{E865 preliminary \kpmm\ data from 1997.}\label{f865_pmm}
\end{figure}

During 1997, due to the very short running time, a special set of triggers were collected at lower intensity. 
Three different decays were studied: \kpmm\ and \kppen\  were collected at low intensity due to trigger rates, 
and \kpen\ were collected at even lower intensity for better control of systematics.
In addition, a new beam detector was installed during the 1997 run, which allows for improved resolution
on decays with missing neutrinos such as \kpen\ and \kppen.
E865  approximately tripled the world sample of \kpmm\ (previously $\sim$200 events
from E787). A preliminary analysis of the 1997 \kpmm\ data sample shown in figure~\ref{f865_pmm} has $\sim$400 events
at the reconstructed $M_{\pi\mu\mu} = M_K$.
In addition, approximately 50,000 \kpen\ events were collected with the aim of improving the understanding of the
CKM matrix element \vus. Analysis of this data should be completed within the next year.
Some 300,000 events of \kppen\ were also collected. This is $\sim\times10$ the previous world sample. This
decay mode is important to \XPT, for which it provides significant input, including information on
low energy $\pi$--$\pi$ scattering.

\begin{figure}[htb]
\center\epsfig{file=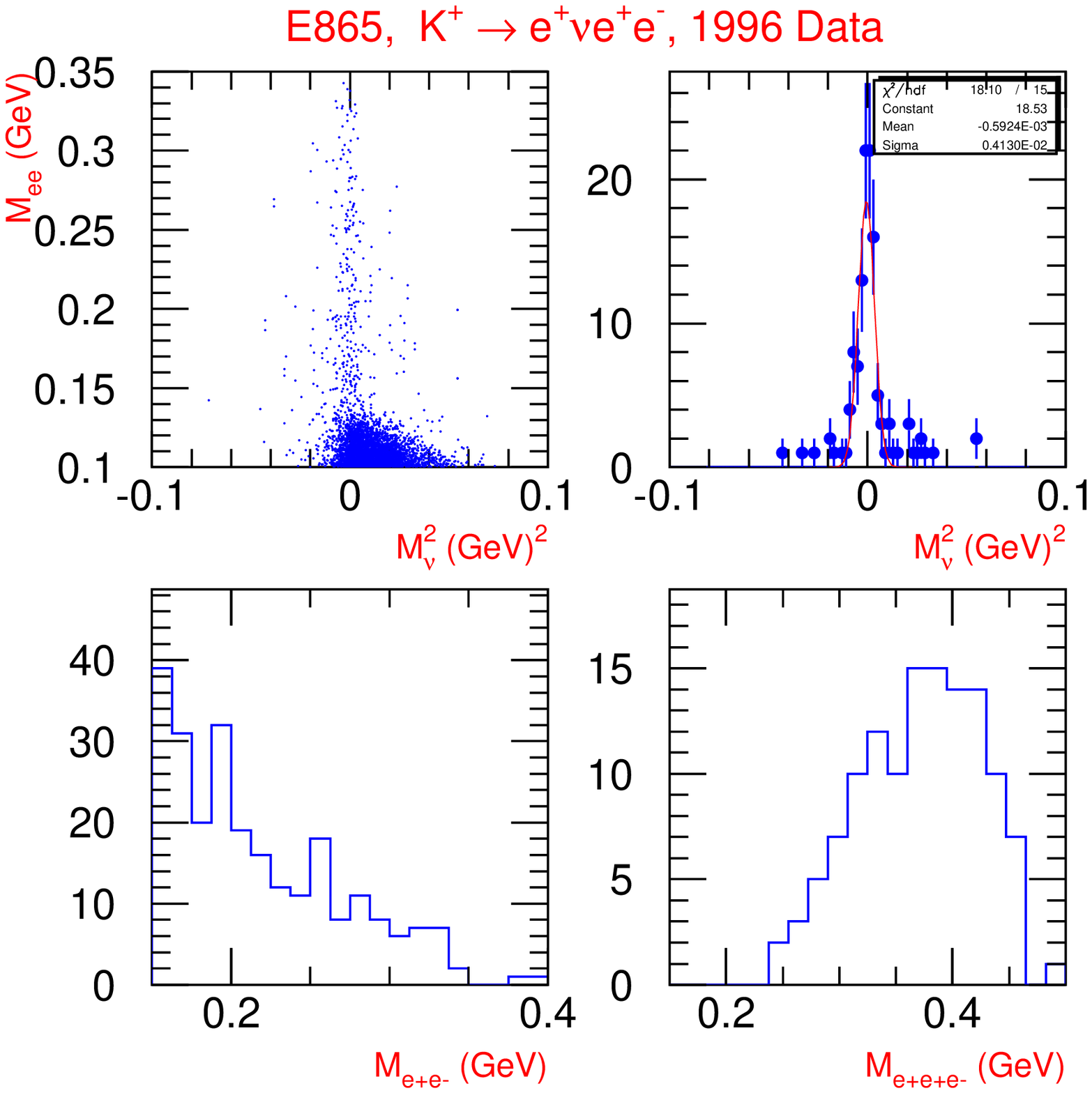,width=5.0in,height=2.0in,angle=0}
\center\epsfig{file=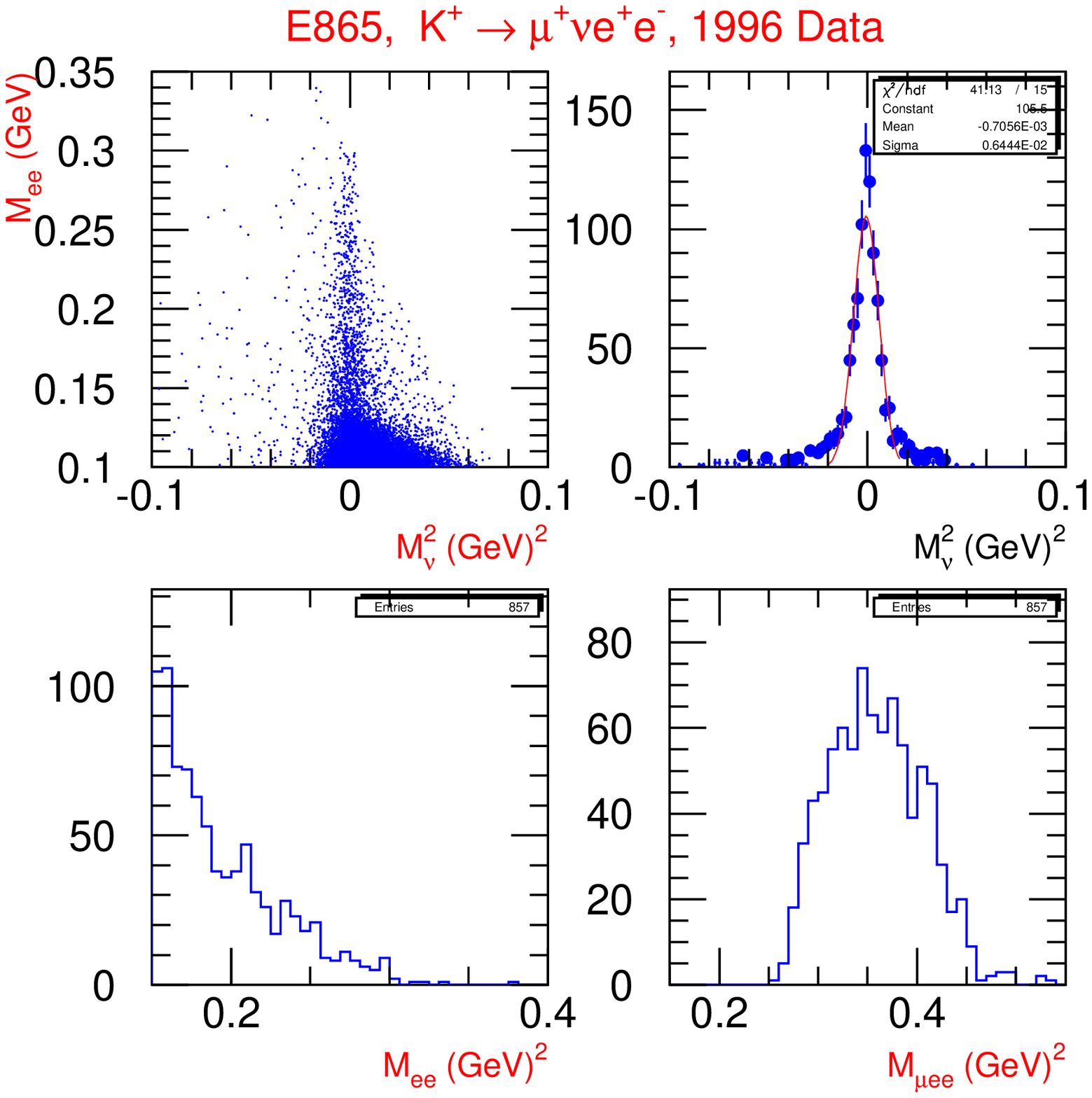,width=5.0in,height=2.0in,angle=0}
\caption{E865 preliminary \kenee\ and \kmnee\ data from 1996.}\label{f865_enee}
\end{figure}

The radiative decays \kenee\ and \kmnee, as well as \kmng, are governed by the form factors $F_A$, $F_V$ and $R$.
These provide a test of \XPT. Preliminary data from the 1996 run for \kenee\ and \kmnee\ are shown 
in figure~\ref{f865_enee}. The $\sim$100 \kenee\ events represents an increase of $\sim$20 over the previous
world sample of 4 events. The $\sim$700 \kmnee\ events increase the world sample by $\sim$50.

\begin{figure}[htb]
\center\epsfig{file=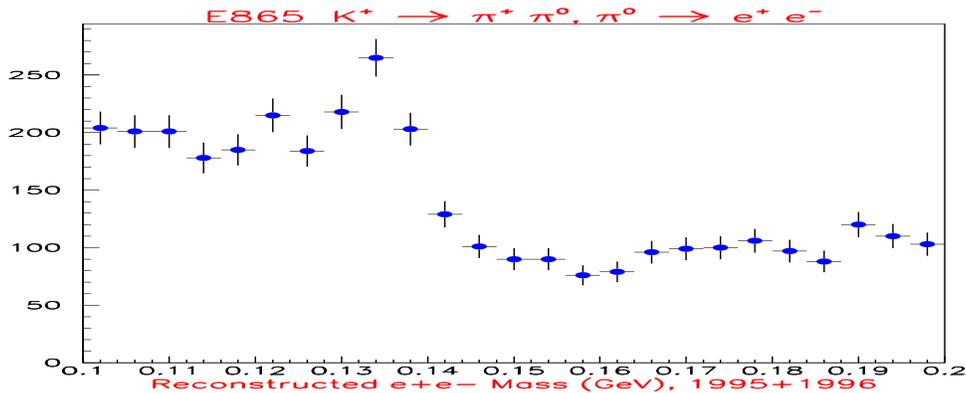,width=5.0in,height=2.0in,angle=0}
\caption{E865 preliminary \pee\ data from 1995 and 1996.}\label{f865_ee}
\end{figure}

Finally, by tagging $\pi^\circ$'s from \kpp\ decays, the decay \pee\ can be studied. From the combined 1995--96 
data set, a preliminary analysis of \pee\ is shown in figure~\ref{f865_ee}.
This decay has had a controversial history, with early experiments measuring a branching ratio
much larger than expected in the SM.~\cite{pi0ee_exp1,pi0ee_exp2} 
The most recent measurement from E851 is consistent with SM predictions
B(\pee) = $(6.9\pm2.3\pm0.6)\times10^{-8}$ (Ref~\citenum{pi0ee}).
 It was based on $\sim$21 events. From the plot in figure~\ref{f865_ee},
it is apparent that E865 has significantly more events.

\subsection{E871}

\begin{figure}[htb]
\hfil\center\epsfig{file=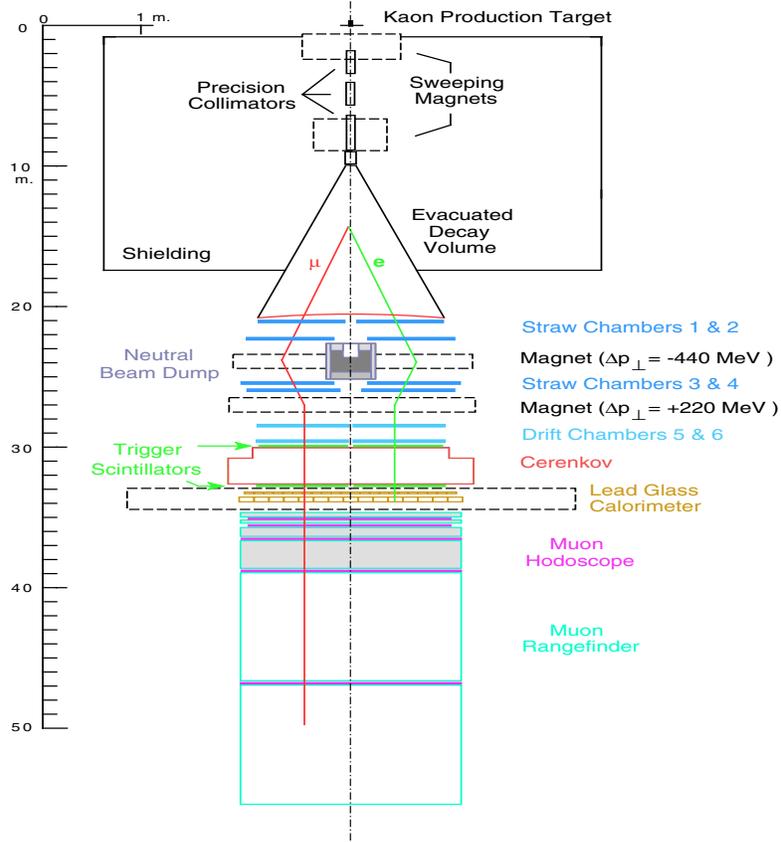,width=5.0in,height=5.0in,angle=0}\hfil
\caption{E871 experimental layout.}\label{f871_exp}
\end{figure}

The goal of the E871 experiment is to push the sensitivity for the lepton flavor violating decay
\klme\ to $10^{-12}$.
The experiment has tried a novel approach to attain this extraordinarily high sensitivity; the very high 
intensity neutral beam is stopped in a tungsten beam plug in the middle of the first 
spectrometer magnet~\cite{plug}. This allows the downstream detectors to operate at relatively low
rates. A layout of the E871 detector is shown in Figure~\ref{f871_exp}.
The neutral kaon beam is produced at 3.75$^\circ$ and
is defined by a series of collimators and sweeping magnets. 
The kaons decay in an 11 m vacuum decay volume that is terminated on the downstream end by a thin Kevlar/Mylar
window. The neutral beam (n/K $\sim$ 20:1) is stopped in the beam plug, whereas the $K_L$ decay products are
tracked in the spectrometer on each side of the beam plug.
 The first magnet imparts a $P_T$ kick of $\sim440$ MeV/c and
the second magnet imparts a $P_T$ kick of $\sim220$ MeV/c in the opposite direction; this leaves two
body decays parallel to the incident beam.

The general design of the E871 experiment is very similar to E791. 
It is a two arm spectrometer with
redundant measurements of the momentum in each arm and redundant $e$ and $\mu$ identification.
Electrons are identified in hydrogen Cherenkov counters and in a PbG array. Muons are identified in an
active iron filter (Fe and scintillator) and in a muon range finder (proportional counters separated by Al or marble
sheets).
One of the primary backgrounds is from \klpen\ where both the $\pi$ and $e$ are mis-identified: in this case the
reconstructed mass can be greater than or equal to $M_K$. This background is eliminated with redundant
lepton identification. The second major background is from \klpen\ decays where the $\pi$ decays to or is misidentified
as a $\mu$. The endpoint of this spectrum is 8 MeV/c$^2$ below $M_K$, so this background can be eliminated
by the high precision, redundant momentum measurements. 
The mass resolution of the spectrometer is $\sigma_{M_{\pi\pi}} \sim 1.1$ MeV/c$^2$.
If the $\pi$ decays to a $\mu$ and one hit is missing in
the x-view in one of the upstream tracking stations, the track can have a good $\chi^2$ and the reconstructed
mass can be greater than or equal to $M_K$. This background is eliminated by adding a third x-measurement to
each tracking station, so that if one hit is missing there are still two hits to constrain the x-position and
by making a tight requirement on the muon range compared to the expectation from the momentum.

  The experiment is designed to take $\times4$ more protons on target than E791 and have a much larger vacuum 
decay tank ($\times3$ larger acceptance). The experiment ran in the B5 beamline with 15--20 Tp on target, 
giving $\sim3\times10^8$ $K_L$/spill and $\sim10^7$ $K_L$ decays per spill.
The L0 trigger rate (requiring parallel tracks) was $\sim$70k and the L1 trigger rate (requiring
the lepton identification) was 10k. The final trigger rate, after mass and transverse
momentum cuts, was $\sim$500/spill.

  The upstream drift chambers in E791 have been replaced with 5 mm diameter straw chambers and are
operated with faster gas (C$_2$H$_6$--CF$_4$), in order to operate efficiently in the high rate
environment near the beam plug.
An additional tracking station has been added to the E791
configuration and all of the chambers have added a third x-measurement. 
The trigger was upgraded to take advantage of the `parallelism' of the tracks from 2-body decays
downstream of the spectrometer magnets. The trigger rate is further reduced by requiring
spatial correlations with the lepton identification systems. The lepton identification systems
have also been more finely segmented.

\begin{figure}[htb]
\center\epsfig{file=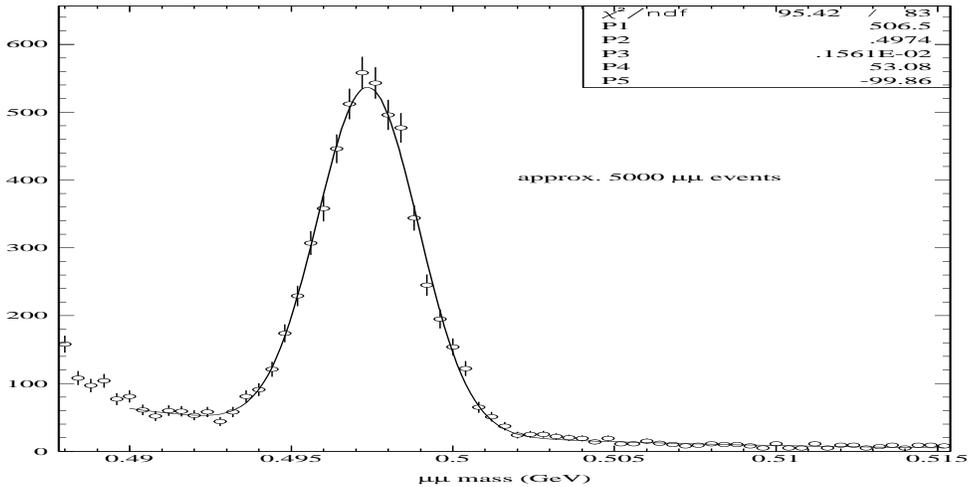,width=5.in,height=2.5in,angle=0}
\caption{Preliminary E871 $\mu\mu$ mass plot.}\label{f871_mm}
\end{figure}

 The E871 experiment has two large data sets from the 1995 and 1996 running periods.
The experiment ran for 25 weeks in 1995 and, after a vacuum window failure forced the rebuilding of
the straw chamber system,  for 16 weeks in 1996.
The 1995--96 data set is being analyzed and results should be forthcoming within the next year.
A preliminary study of the \klmm\ decay shows $\sim$5000 events, implying 
 a sensitivity of $\sim1.5\times10^{-12}$, close to the design goal. A plot of
the \klmm\ invariant mass is shown in figure~\ref{f871_mm}.
This is $\stackrel{>}{\sim}\times7$ more than from E791. With the increased sensitivity E871 should `close the window'
on new physics in \klee. In the SM the branching ratio is a few $\times10^{-12}$, so E871 should observe several
\klee\ events.

%\vfill
%\eject
\section{AGS--2000}

The AGS will become the injector for RHIC (Relativistic Heavy Ion Collider) beginning in 1999.
The AGS will be needed for injection up to twice a day for $\sim$2 hours; the remaining 20 hours
 will be available for running
fixed target experiments with either heavy ions or protons. 
The AGS will remain the most intense source of protons with energy above the kaon production threshold.
In particular the AGS will remain the most intense source of low energy kaons.
The marginal cost for the extra running time
will be substantially less than the current cost of the program. Several kaon decay experiments are being
considered or are already proposed for running in the AGS--2000 era.

\subsection{\kpnn}

E787 has seen one event with B(\kpnn)=4.2$^{+9.7}_{-3.5}\times10^{-10}$. Several improvements
have already been made or are underway, including: lowering the kaon momentum (increasing the number of 
stopped kaons, while
keeping the rates in the detector low), increasing the duty factor of the AGS (increasing sensitivity
without increasing rates), reduced deadtime through trigger and DAQ upgrades, additional transient digitizers
on the beam elements, finer segmentation of the beam chamber, and an additional layer of Pb-scintillator for the
central photon veto. Additional improvements were considered at the AGS--2000 Workshop. The goal
of the \kpnn\ working group was to design an experiment to measure \vtd\ to $\stackrel{<}{\sim}$15\%. The working
group concluded that the optimal strategy would be to build on the vast experience and success of
the E787 experiment and that,
with modest upgrades to the E787 experiment and beamline, achieving this goal should be possible.~\cite{pnn2000}

\subsection{E926}

The decay \klpnn, the so called `Golden Mode', is a purely CP-violating decay that is entirely dominated by 
short distance physics
involving the top quark. As in \kpnn\ the hadronic matrix element can be derived from 
\kpen. Due to the small size of the imaginary part of the CKM matrix elements, the charm quark contribution is negligible.
The remaining theoretical uncertainty is even  smaller than that for the charged mode, [$O$(1\%)].~\cite{bf}
The branching ratio is given by
\begin{equation}
B(K_L \to \pi^0 \nu\bar\nu) \approx 4.3 \times 10^{-10} \eta^2 A^4 \sim  2 \times10^{-11}
\end{equation}

\begin{figure}[htb]
\center\epsfig{file=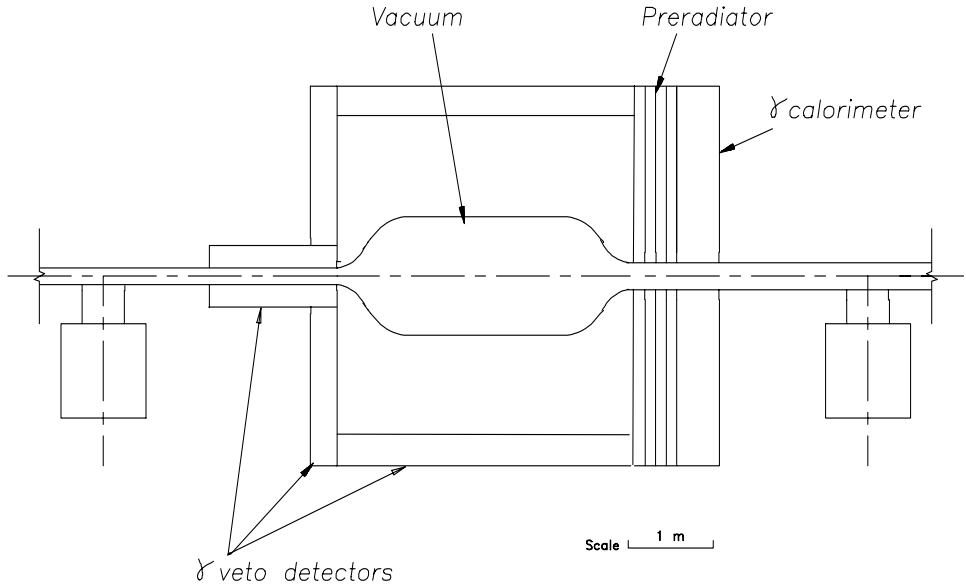,width=5.0in,height=3.0in,angle=0}
\caption{E926 experimental layout.}\label{f926_det}
\end{figure}

The experimental challenges are formidable, not only because of the small expected branching ratio.
The experimental signature is even weaker than in the charged mode. The E926~\cite{e926} experiment has chosen
to measure as many kinematic variables as possible: the direction, time, position and energies of the photons from the $\pi^\circ$
and the velocity of the $K_L$. The velocity of the $K_L$ will be determined from the arrival times of the photons,
the reconstructed decay vertex and the time of the $K_L$ creation. This last quantity will be determined to
$\sim$200ps from the use of a bunched beam from the AGS. The reconstruction of the $\gamma$ directions will be
achieved with a 1.5 $X_0$ preconverter (scintillator and chambers). The energy will be measured with a Pb
and scintillating fiber calorimeter similar to the KLOE design~\cite{kloe}, but with a much enhanced scintillator to Pb ratio. 
The rest of the decay volume will be
surrounded by photon veto systems. A diagram of the proposed detector is shown in figure~\ref{f926_det}.

To maximize the $K_L$ momentum resolution and to reduce backgrounds from neutron interactions the
experiment will run at a large production angle ($\sim45^\circ$). The $K_L$ beam will be a flat beam,
125 mr by 4 mr, which provides an additional kinematic constraint ($y$ of the vertex). The neutral beam and decay
volume will be at moderately high vacuum ($10^{-7}$ torr) to suppress neutron backgrounds. The
largest remaining background is from \klpp\ which, while also CP-violating, is $10^{8}$ times
larger. The E926 design goal is a $\pi^\circ$ veto inefficiency of $10^{-8}$. Additional suppression
of the \klpp\ of a factor of 50 can be achieved with a kinematic cut on the $\pi^\circ$ momentum
in the $K_L$ center of mass system at about 190 MeV/c (35\% acceptance for \klpnn).

The experiment plans to run with 50 Tp on target, giving $\sim2.5\times10^8 K_L$'s and $\sim2\times10^{7} K_L$
decays per pulse. With an 8000 hour run,  $\sim$70 \klpnn\ decays should be detected, with a background
of $\sim$7 events (almost entirely from \klpp). 

Measurements of the \kpnn\ and \klpnn\ branching ratio may provide the most precise determination
of the CKM parameters $\rho$ and $\eta$. In any event, a determination of these parameters in the
kaon system will make for a valuable comparison with values obtained for the B system from the
B-factories.

\subsection{E923}

  E923~\cite{e923} will search for a T-violating polarization of the $\mu^+$ normal to the decay 
plane in the decay \kmuiii. The
current limits on this process, $-0.009<P_\mu^T<0.007$(95\%CL), 
come from earlier BNL experiments E696 and E735~\cite{Akm3_0,Akm3_1,Akm3_2,Akm3_3}
which collected data from 1978--80.
Although the SM does not predict T-violation in \kmuiii, CP-violation is not well understood and, for example,  some
proposed extensions to the SM designed to explain the baryon asymmetry in the universe also predict
transverse muon polarization.~\cite{garisto,belanger}

\begin{figure}[htb]
\center\epsfig{file=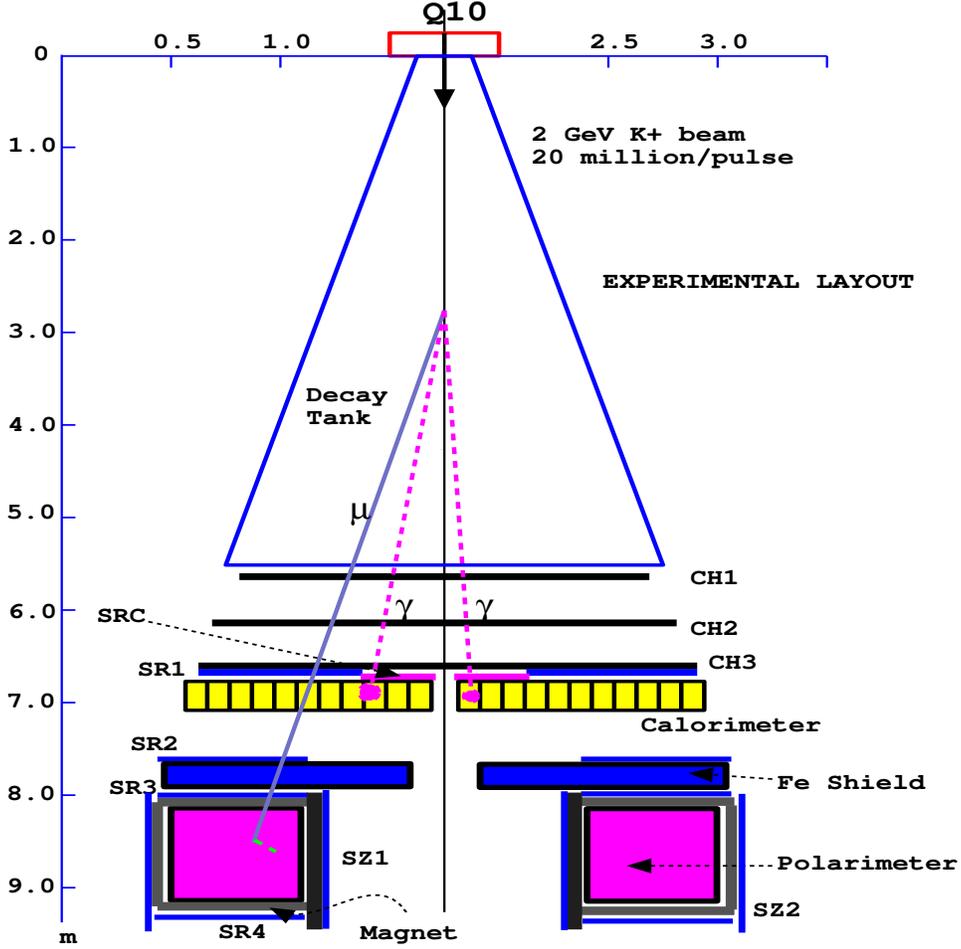,width=5.0in,height=5.0in,angle=-90}
\caption{E923 experimental layout.}\label{f923_det}
\end{figure}

The earlier experiment collected $2.1\times10^7 K^+$ ($1.2\times10^7 K_L$) decays with an unseparated 4 GeV/c
$K^+$ beam.
The new experiment will use a separated (K$^+$:$\pi^+$ = 1/1.2) 2 GeV beam with 2$\times10^{7}$ K$^+$/pulse 
(30 Tp protons on target) and 6$\times10^6$ decays/pulse. 
A drawing of the detector is shown in figure~\ref{f923_det}.
The new detector has a larger acceptance and achieves better background rejection. The background rejection
is improved by fully reconstructing the event with the large acceptance,  finely-segmented calorimeter 
and the tracking chambers Ch1--3.
The $\gamma$'s from the $\pi^\circ$'s are detected in an electromagnetic calorimeter (Pb-scintillator ---
Shashlyk design). The $\mu^+$'s are tracked into the graphite polarimeter. The polarimeter has 96 graphite
wedges, with chambers between each wedge. An asymmetry in the rate of clockwise vs counter-clockwise
muon decays would signal a possible T-violating muon polarization.
An axial magnetic field (along the beam direction) of $\sim70$ G is applied to the polarimeter, with the
field direction reversed every spill, allowing for the cancellation of many possible systematic errors.

  With a total analyzing power of 23\% and a 2000 hour run, the statistical sensitivity will reach $1.3\times10^{-4}$.

The same apparatus will be used to measure the T-violating polarization in 
\kmng\ events.  Such a measurement is sensitive to 
non-standard pseudo-scalar as well as vector interactions and is 
therefore complimentary to the measurement in \kmuiii\ decays.~\cite{geng94} 
Studies are in progress to ascertain if a sensitivity of  0.001 can be achieved, at which point
final state interactions may produce a non T-violating 
polarization out of the decay plane.

%In addition this experiment will look a T-violating polarization of muons out of the decay plane for \kmng.

\subsection{E927}

The E927~\cite{e927} experiment proposes to measure the CKM matrix element \vus.
The goal is to measure the \kpen\ branching ratio to 0.7\%. 
A $K^+$ beam will be stopped in a scintillating fiber target. The outgoing positron will
be tracked in a drift chamber and identified in plastic scintillator and a plexiglass
Cherenkov counter. The $\pi^\circ$ photons and $e^+$ energy will be measured in the Crystal Ball
detector.
The estimate for \kpen\ acceptance is as high as 98.5\% with a 1.3\% contamination from predominantly 
\kpp\ with a small \kpmn\ component.

\section{Conclusions}

The AGS kaon program has long and rich history. The most rare particle decay yet has been observed in \kpnn.
The most sensitive kaon decay search comes from \klme. A number of interesting new processes have been observed. Further
improvements can be expected from the current experiments. Future extensions of the program should provide a significant 
determination of the CKM matrix element \vtd\ and the CP-violating parameter $\eta$. Significantly improved limits on, 
or perhaps observation of T-violation in the decay \kpmn\ should be expected.
Improvements in the knowledge of the CKM matrix element \vus\ should also be forthcoming.

\section{Acknowledgements}
  I would like to thank Laurie Littenberg, Milind Diwan, Bill Molzon, Hong Ma, Mike Zeller, 
Bill Morse, Phil Pile, Tom Roser, I.-H. Chiang, Brad Tippens,  Robin Appel and Mike Hebert for comments
and useful discussions regarding this paper and for access to the data presented here.

\section{Appendix}

Publications from the recent AGS program are listed in tables~\ref{ags_pub1}--\ref{ags_pub4}.
\begin{table}[htbp]
\begin{center}
\begin{tabular}{|l|}\hline
\multicolumn{1}{|c|}{E696/E735/E749/E780/E845} \\ \hline\hline
M.P.~Schmidt {\it et al.}, Phys.Rev.Lett. {\bf 43}, 556--560 (1979).	\\ \hline
W.M.~Morse {\it et al.}, Phys.Rev. {\bf D21}, 1750--1766 (1980).	\\ \hline
M.K.~Campbell {\it et al.}, Phys.Rev.Lett. {\bf 47}, 1032--1035 (1981).	\\ \hline
S.R.~Blatt {\it et al.}, Phys.Rev. {\bf D27}, 1056--1068 (1983).	\\ \hline
J.K.~Black {\it et al.}, Phys.Rev.Lett. {\bf 54}, 1628--1630 (1985).	\\ \hline
H.B.~Greenlee {\it et al.}, Phys.Rev.Lett. {\bf 60}, 893--896 (1988).	\\ \hline
E.~Jastrzembski {\it et al.}, Phys.Rev.Lett. {\bf 61}, 2300--2303 (1988).	\\ \hline
S.F.~Schaffner {\it et al.}, Phys.Rev. {\bf D39}, 990--993 (1989).	\\ \hline
H.B.~Greenlee, Phys.Rev. {\bf D42}, 3724--3731 (1990).	\\ \hline
K.E.~Ohl {\it et al.}, Phys.Rev.Lett. {\bf 64}, 2755--2758 (1990).	\\ \hline
W.M.~Morse {\it et al.}, Nucl.Phys. {\bf A527}, 717--720 (1991).	\\ \hline
K.E.~Ohl {\it et al.}, Phys.Rev.Lett. {\bf 65}, 1407--1410 (1990).	\\ \hline
W.M.~Morse {\it et al.}, Phys.Rev. {\bf D45}, 36--41 (1992).	\\ \hline
M.R.~Vagins {\it et al.}, Phys.Rev.Lett. {\bf 71}, 35--37 (1993).	\\ \hline
\end{tabular}
\caption{Publications from AGS experiment E780/E845.}\label{ags_pub1}
\end{center}
\end{table}
\begin{table}[htbp]
\begin{center}
\begin{tabular}{|l|}\hline
\multicolumn{1}{|c|}{E777/E851/E865} \\ \hline\hline
N.J.~Baker {\it et al.}, Phys.Rev.Lett. {\bf 59}, 2832--2835 (1987).	\\ \hline
C.~Campagnari {\it et al.}, Phys.Rev.Lett. {\bf 61}, 2062--2065 (1988).	\\ \hline
A.M.~Lee {\it et al.}, Phys.Rev.Lett. {\bf 64}, 165--168 (1990).	\\ \hline
C.~Alliegro {\it et al.}, Phys.Rev.Lett. {\bf 68}, 278--281 (1992).	\\ \hline
A.~Deshpande {\it et al.}, Phys.Rev.Lett. {\bf 71}, 27--30 (1993).	\\ \hline
\end{tabular}
\caption{Publications from AGS experiment E777/E851/E865.}\label{ags_pub2}
\end{center}
\end{table}
\begin{table}[htbp]
\begin{center}
\begin{tabular}{|l|}\hline
\multicolumn{1}{|c|}{E791/E871} \\ \hline\hline
R.D.~Cousins {\it et al.}, Phys.Rev. {\bf D38}, 2914--2917 (1988).	\\ \hline
C.~Mathiazhagan {\it et al.}, Phys.Rev.Lett. {\bf 63}, 2181--2184 (1989).	\\ \hline
C.~Mathiazhagan {\it et al.}, Phys.Rev.Lett. {\bf 63}, 2185--2188 (1989).	\\ \hline
A.P.~Heinson {\it et al.}, Phys.Rev. {\bf D44}, R1--R5 (1991).	\\ \hline
K.~Arisaka {\it et al.}, Phys.Rev.Lett. {\bf 70}, 1049--1052 (1993).	\\ \hline
A.P.~Arisaka {\it et al.}, Phys.Rev.Lett. {\bf 71}, 3910--3913 (1993).	\\ \hline
A.P.~Heinson {\it et al.}, Phys.Rev. {\bf D51}, 985--1013 (1995).	\\ \hline
J.~Frank {\it et al.}, IEEE Trans.Nucl.Sci. {\bf NS--36}, 79--85 (1989).	\\ \hline
C.J.~Kenney {\it et al.}, IEEE Trans.Nucl.Sci. {\bf NS--36}, 74--78 (1989).	\\ \hline
D.M.~Lee {\it et al.}, NIM {\bf A256}, 329--332 (1987).	\\ \hline
R.D.~Cousins {\it et al.}, IEEE Trans.Nucl.Sci. {\bf NS--36}, 646--649 (1989).	\\ \hline
K.A.~Biery {\it et al.}, IEEE Trans.Nucl.Sci. {\bf NS--36}, 650--652 (1989).	\\ \hline
R.D.~Cousins {\it et al.}, NIM {\bf A277}, 517 (1989).	\\ \hline
\end{tabular}
\caption{Publications from AGS experiment E791/E871.}\label{ags_pub3}
\end{center}
\end{table}
\begin{table}[htbp]
\begin{center}
\begin{tabular}{|l|}\hline
\multicolumn{1}{|c|}{E787} \\ \hline\hline
M.S.~Atiya {\it et al.}, Phys.Rev.Lett. {\bf 63}, 2177--2180 (1989).	\\ \hline
M.S.~Atiya {\it et al.}, Phys.Rev.Lett. {\bf 64}, 21--24 (1990). 	\\ \hline
M.S.~Atiya {\it et al.}, Phys.Rev.Lett. {\bf 65}, 1188--1191 (1990).	\\ \hline
M.S.~Atiya {\it et al.}, Phys.Rev.Lett. {\bf 66}, 2189--2192 (1991).	\\ \hline
M.S.~Atiya {\it et al.}, Phys.Rev.Lett. {\bf 69}, 733--736 (1992).	\\ \hline
M.S.~Atiya {\it et al.}, Nucl.Phys. {\bf A527}, 727c--729c (1991).	\\ \hline
M.S.~Atiya {\it et al.}, Phys.Rev.Lett. {\bf 70}, 2521--2524 (1993).	\\ \hline
M.S.~Atiya {\it et al.}, Phys.Rev. {\bf D48}, 1--4 (1993).	\\ \hline
S.~Adler {\it et al.}, Phys.Rev.Lett. {\bf 76}, 1421--1424 (1996).	\\ \hline
S.~Adler {\it et al.}, Phys.Rev.Lett. {\bf 79}, 2204--2207 (1997).	\\ \hline
P.~Kitching {\it et al.}, Phys.Rev.Lett. {\bf 79}, 4079--4082 (1997).	\\ \hline
S.~Adler {\it et al.}, Phys.Rev.Lett. {\bf 79}, 4756--4759 (1997).	\\ \hline
M.S.~Atiya {\it et al.}, IEEE Trans.Nucl.Sci. {\bf NS--36}, 813--817 (1989). \\ \hline
M.S.~Atiya {\it et al.}, NIM {\bf A279}, 180--185 (1989). \\ \hline
M.S.~Atiya {\it et al.}, NIM {\bf A321}, 129--151 (1992).	\\ \hline
M.~Burke  {\it et al.}, IEEE Trans.Nucl.Sci. {\bf NS--41}, 131 (1994).	\\ \hline
M.~Kobayashi {\it et al.}, NIM {\bf A337}, 355--361 (1994).	\\ \hline
I.-H.~Chiang {\it et al.}, IEEE Trans.Nucl.Sci. {\bf NS--42}, 394--400 (1995). \\ \hline
D.A.~Bryman {\it et al.}, NIM {\bf A396}, 394--404 (1997). 	\\ \hline
T.K.~Komatsubara {\it et al.}, accepted for publication in NIM {\bf A}. \\ \hline
E.W.~Blackmore {\it et al.}, accepted for publication in NIM {\bf A}.	\\ \hline
\end{tabular}
\caption{Publications from AGS experiment E787.}\label{ags_pub4}
\end{center}
\end{table}

\end{document}